\documentclass[twocolumn]{aastex701}

\usepackage{natbib}
\setcitestyle{aysep={}}
\usepackage{graphicx}	
\usepackage{amsmath}	
\usepackage{amssymb}	
\usepackage{subfigure}
\usepackage{threeparttable}
\usepackage{balance}
\usepackage{indentfirst}
\usepackage{CJKutf8}
\usepackage{siunitx}
\usepackage{array}
\usepackage{booktabs}
\usepackage{rotating}

\newcommand{\ffm}{f_\text{F,mean}}
\newcommand{\ffr}{f_\text{F,rms}}
\newcommand{\fsm}{f_{\sigma,\text{mean}}}
\newcommand{\fsr}{f_{\sigma,\text{rms}}}
\newcommand{\Fm}{\text{FWHM}_\text{mean}}
\newcommand{\Fr}{\text{FWHM}_\text{rms}}
\newcommand{\sm}{\sigma_\text{mean}}
\newcommand{\sr}{\sigma_\text{rms}}

\newcommand{\dhbm}{D_{\text{H}\beta,\text{mean}}}

\newcommand{\thb}{\tau_{\text{H}\beta}}
\newcommand{\lv}{\ifmmode L_{5100} \else $L_{5100}$\ \fi}
\newcommand{\kms}{\ifmmode {\rm km\ s}^{-1} \else km s$^{-1}$\ \fi}
\newcommand{\ergs}{\ifmmode {\rm erg\ s}^{-1} \else erg s$^{-1}$\ \fi}
\newcommand{\ergc}{\ifmmode {\rm erg\ s^{-1}cm^{-2}} \else $\rm erg\ s^{-1} cm^{-2}$\ \fi}
\newcommand{\lb}{\ifmmode L_{\rm Bol} \else $L_{\rm Bol}$\ \fi}
\newcommand{\ledd}{\ifmmode L_{\rm Edd} \else $L_{\rm Edd}$\ \fi}
\newcommand{\hb}{\ifmmode H\beta \else H$\beta$\ \fi}
\newcommand{\ha}{\ifmmode H\alpha \else H$\alpha$\ \fi}
\newcommand{\oiii}{\ifmmode \rm [O\ \sc{III}] \else $\rm [O\ \sc{III}]$\ \fi}
\newcommand{\nii}{\ifmmode \rm [N\ \sc{II}] \else $\rm [N\ \sc{II}]$\ \fi}
\newcommand{\mgii}{\ifmmode \rm Mg\ \sc{II} \else $\rm Mg\ \sc{II}$\ \fi}
\newcommand{\civ}{\ifmmode \rm C\ \sc{IV} \else $\rm C\ \sc{IV}$\ \fi}
\newcommand{\ciii}{\ifmmode \rm C\ \sc{III]} \else $\rm C\ \sc{III]}$\ \fi}
\newcommand{\siiv}{\ifmmode \rm Si\ \sc{IV} \else $\rm Si\ \sc{IV}$\ \fi}
\newcommand{\mbh}{\ifmmode M_{\rm BH}  \else $M_{\rm BH}$\ \fi}
\newcommand{\msun}{M_{\odot}}
\newcommand{\rfe}{\ifmmode R_{\rm Fe} \else $R_{\rm Fe}$\ \fi}
\newcommand{\sst}{\ifmmode \sigma_{\rm \ast}\else $\sigma_{\rm \ast}$\ \fi}
\newcommand{\dhb}{\ifmmode D_{\rm H\beta} \else $D_{\rm H\beta}$\ \fi}
\newcommand{\leddR}{\ifmmode L_{\rm Bol}/L_{\rm Edd}\else $L_{\rm Bol}/L_{\rm Edd}$\ \fi}
\newcommand{\feii}{\ifmmode \rm Fe\ \sc{II} \else $\rm Fe\ \sc{II}$\ \fi}
\newcommand{\mdot}{\ifmmode \dot{\mathscr{M}}  \else $\dot{\mathscr{M}}$\ \fi}
\newcommand{\rhb}{\ifmmode R_{\rm BLR}({\rm H\beta})  \else $R_{\rm BLR}({\rm H\beta})$ \ \fi}
\newcommand{\shb}{\ifmmode \sigma_{\rm H\beta} \else $\sigma_{\rm \hb}$\ \fi}
\newcommand{\RL}{\ifmmode R_{\rm BLR}({\rm H\beta}) - L_{\rm 5100} \else $R_{\rm BLR}({\rm H\beta}) - L_{\rm 5100}$ \ \fi}
\newcommand{\ms}{\ifmmode M_{\rm BH}-\sigma_{\ast} \else $M_{\rm BH}-\sigma_{\ast}$\ \fi}
\newcommand{\bd}{\ifmmode \rm \ha/\hb \else $\rm \ha/\hb$ \fi}

\begin{document}

\title{The virial factor $f$ of the \hb Broad-line for NGC 5548 and NGC 4151}

\author[orcid=0009-0008-3600-670X]{Shao-Jun Li}
\affiliation{School of Physics and Technology, Nanjing Normal University, Nanjing 210023, People's Republic of China}
\email{whbian@njnu.edu.cn}
\author{Xiang-Wei Ning}
\affiliation{School of Physics and Technology, Nanjing Normal University, Nanjing 210023, People's Republic of China}
\email{whbian@njnu.edu.cn}
\author[orcid=0009-0008-5295-3977]{Yan-Song Ma}
\affiliation{School of Physics and Technology, Nanjing Normal University, Nanjing 210023, People's Republic of China}
\email{whbian@njnu.edu.cn}
\author{Yi Tang}
\affiliation{School of Physics and Technology, Nanjing Normal University, Nanjing 210023, People's Republic of China}
\email{whbian@njnu.edu.cn}
\author[orcid=0000-0002-2121-8960]{Wei-Hao Bian}
\affiliation{School of Physics and Technology, Nanjing Normal University, Nanjing 210023, People's Republic of China}
\email[show]{whbian@njnu.edu.cn}

\begin{abstract}
The variability of virial factor $f$  is investigated for two active galactic nucleus, NGC 5548 and NGC 4151,  which had been previously reverberation mapped (RM) over  20 times in the past 30 years. Using four velocity tracers from the broad \hb width at half-maximum ($\rm FWHM_{\hb} $) or the line dispersion from the mean or rms spectra,  $f$ for each RM epoch are calculated.
Significant correlations are found between $f$ and observational parameters, such as  the broad line widths, the Eddington ratios and the line profile shapes.
For NGC 5548, $f \propto \rm {FWHM}_{mean}^{-0.70\pm0.13}$ and for NGC 4151, $f \propto \rm {FWHM}_{mean}^{-3.31\pm0.59}$.
This suggests that a variable $f$ should be included to weight the virial SMBH mass.
Using a simple model of thick-disc broad line regions (BLRs), we show that changes in mean inclination can explain $f$ variation. The inclination range is  $14.1-40.6$ deg for NGC 5548 and $14.0-55.1$ deg for NGC 4151.
Modeling  the light curves of $f$ with a damped random walk process yields  mean timescales of 638 and 668 days, consistent with BLR dynamical timescales within uncertainties. This indicates that $f$ variations  are linked to BLR dynamics, likely due to changes in geometry or inclination.

\end{abstract}

\section{Introduction} \label{sec:1}
Active galactic nucleus (AGNs) can be classified into Type I and Type II, depending on whether broad line regions (BLRs) can be viewed directly \citep[e.g.,][]{Antonucci_1985, Netzer_2013}. For Type I AGNs,  the BLR clouds emitting broad lines can be used as a probe of the gravitational potential of supermassive black holes (SMBHs), of which the virial mass can be derived as follows \citep[e.g.,][]{Peterson_2004}:
\begin{equation}
    \mbh=f\times\frac{R_{\rm BLR}\times(\Delta V)^{2}}{G} \equiv f \times VP
	\label{eq:1}
\end{equation}
where $G$ is the gravitational constant, \mbh is the black hole mass, $\Delta V$ is the velocity of the BLR clouds and $R_{\rm BLR}$ is the distance from the SMBH to the BLRs. VP is the so-called virial product, $VP \equiv R_{\rm BLR}\Delta V^{2}/G$. $f$ is the virial factor, which is used to characterize the kinematics, geometry, and inclination of the BLR clouds \citep[e.g.,][]{Colin_2006, Yu_2020}.

Considering the photon-ionization model of BLRs, $R_{\rm BLR}$  can be estimated from the reverberation mapping (RM) method \citep[RM,][]{Blandford_1982, Peterson_1993}. The variations in the continuum would lead to the changes in the broad-line emission delayed by a lag time $\tau =(1+z)~ R_{\rm BLR}/c$ for an AGN with redshift  $z$ . The \hb time lag was successfully measured by this RM method for more than 120 AGNs \citep[e.g.,][]{Peterson_2004, Bentz_2013, Wang_2014, Grier_2017b, De_rosa_2018, Shen_2019, Yu_2020, Hu_2021,Lu_2022, Chen_2023} .  
Velocity of the BLR clouds  $\Delta V$ is usually traced by  the full width at half-maximum (FWHM) or the line dispersion ($\sigma_{\rm line}$) of the broad \hb line measured from the mean or rms spectrum \citep{Peterson_2004,Ho_2014,Yu_2020,Lu_2022}.  
For other un-RM AGNs, an empirical \RL relation ($L_{\rm 5100}$ is the continuum luminosity at 5100~\AA) was investigated \citep[e.g.,][]{Kaspi_2000,Bentz_2013, Du_2019, Yu_2020} .
The virial factor $f$ and the empirical \RL relation provide the foundation of \mbh calculation from the single-epoch spectra of  type I AGNs in large spectral surveys \citep[e.g.,][]{Bian_2004, Vestergaard_2006, Shen_2011}

The calibration of $f$ is usually done through other independent methods to derive the SMBH masses in RM AGNs, such as the $\mbh-\sigma_*,~ \mbh-L_{\rm bulge}$ relations for quiescent galaxies, when the measured bulge stellar velocity dispersion $\sigma_*$ or the bulge luminosity $L_{\rm bulge}$ are available  for these RM AGNs \citep[e.g.,][]{Onken_2004, Woo_2015, Ho_2014, Yu_2019,Yang_2024}. 
Comparing the scatter of different VP based on  $\sigma_{\rm \hb}$ or \hb FWHM measured from the rms or mean spectra, \cite{Peterson_2004} preferred the VP based on $\sigma_{\rm \hb}$ from the rms spectrum to calculate the virial $M_{\rm BH}$, and $f$ was therefore assumed as a constant. However, $f$ may be different for each object. Based on the $\mbh-\sigma_*$ relation, for a sample of 34 RM AGNs with measured $\sigma_*$ \citep{Ho_2013,Yu_2019}, it was found that the value of $f$ has a range of about two orders of magnitude and FWHM-based $f$ has a strong relation with $\rm FWHM_{\rm mean}$, e.g. $f_{\rm F,mean} \propto \rm FWHM_{\rm mean}^{-1.11\pm0.27}$. 
Based on \mbh derived from the $\mbh-L_{\rm bulge}$ relation for a sample of 40 RM AGNs with measured $L_{\rm bulge}$, similar correlation was also found with $f_{\rm F,mean} \propto \rm FWHM_{\rm mean}^{-1.42\pm0.26}$ \citep{Yang_2024}. 

In comparison with the sample of RM AGNs, $f$ of each RM epoch for a single multi-RM AGN can be used to investigate $f$ variability and its relation with \hb FWHM, where \mbh is nearly constant. \citet{Pancoast_2011} has introduced the  Code for AGN Reverberation and Modeling of Emission Lines (CARAMEL) to fit self-consistent models of the BLR geometry and dynamics to the RM data sets \citep{Bentz_2009,Denny_2010,Grier_2013}.  
CARAMEL provides a phenomenological description of the BLR dynamics, and thereby the inference of the BLR parameters and associated uncertainties in RM data sets \citep{Pancoast_2014a,Winkel_2025}. 
Up to now, this method provided precise and independent $M_{\rm BH}$ measurements for a sample of 30 objects, including NGC 5548 and NGC 4151 \citep{Pancoast_2014b,Grier_2017a,Bentz_2022,Villafana_2023}. 

NGC 5548 ($z$ = 0.01717) and NGC 4151 ($z$ = 0.003326) are nearby Seyfert galaxies hosting a classical bulge \citep{Ho_2014}, both had been spectroscopy monitored more than 20 observing seasons before 2024 with extreme variability in the past four decades  \citep{Li_2016, Lu_2016, Lu_2022, Chen_2023, Feng_2024}. 
In this paper, the virial factors $f$ of each RM epoch for these two multi-RM AGNs are calculated to investigate the variability of $f$ and its relations with other observational parameters.  This paper is organized as follows. Section \ref{sec:2} presents the data we used. Section \ref{sec:3} is data analysis and Section \ref{sec:4} is discussion. Section \ref{sec:5} summarizes our results. All the cosmological calculations in this paper are assumed as $\Omega_{\Lambda}$ = 0.7, $\Omega_\text{M}$ = 0.3, and $\text{H}_{0}$ = 72 km s$^{-1}$ Mpc$^{-1}$.

\section{Data} \label{sec:2}

\begin{deluxetable}{ccccccccccc}
  \renewcommand\arraystretch{1.382}
  \tablecolumns{11}
  \tabletypesize{\scriptsize}
  \setlength{\tabcolsep}{4.5pt}
  \tablewidth{4.5pt}
  \tablecaption{Properties of NGC 5548 with measured $\tau_\text{cent}$ and velocities of the broad H${\beta}$ line for calibration.\label{tab:table1}}
  \tablehead{
\colhead{Data set}    &
\colhead{Duration}    &
\colhead{log$L_{\text{5100}}$}\textsuperscript{a}    &
\colhead{$\tau_{\text{H}\beta}$}    &
\colhead{FWHM$_\text{mean}$}    &
\colhead{FWHM$_\text{rms}$}    &
\colhead{$\sigma_\text{mean}$}    &
\colhead{$\sigma_\text{rms}$}    &
\colhead{$\lambda_\text{Edd}$}    &
\colhead{VP$_{\sigma,\text{r}}$}  &
\colhead{Ref}\\
\colhead{}    &
\colhead{}    &
\colhead{log(erg s$^{-1}$)}    &
\colhead{(days)}    &
\colhead{(km~s$^{-1}$)}    &
\colhead{(km~s$^{-1}$)}    &
\colhead{(km~s$^{-1}$)}    &
\colhead{(km~s$^{-1}$)}    &
\colhead{}    &
\colhead{($10^{7}M_{\odot}$)}    & 
\colhead{}  \\ 
\colhead{(1)}    &
\colhead{(2)}    &
\colhead{(3)}    &
\colhead{(4)}    &
\colhead{(5)}    &
\colhead{(6)}    &
\colhead{(7)}    &
\colhead{(8)}    &
\colhead{(9)}    &
\colhead{(10)}   &
\colhead{(11)}
}
\startdata
    Year 1 & 1988 Dec-1989 Oct & 43.30$\pm$0.09 & 19.70$^{+1.50}_{-1.50}$ & 4674$\pm$63 & 4044$\pm$199 & 1934$\pm$5 & 1687$\pm$56 & 0.030 & 1.09$^{+0.11}_{-0.11}$ & 1,2,3 \\
    Year 2 & 1989 Dec-1990 Oct & 43.03$\pm$0.13 & 18.60$^{+2.30}_{-2.10}$ & 5418$\pm$107 & 4664$\pm$324 & 2223$\pm$20 & 1882$\pm$83 & 0.017 & 1.29$^{+0.20}_{-0.18}$ &1,2,3 \\
    Year 3 & 1990 Nov-1991 Oct & 43.23$\pm$0.08 & 15.90$^{+2.50}_{-2.90}$ & 5236$\pm$87 & 5776$\pm$237 & 2205$\pm$16 & 2075$\pm$81 & 0.026 & 1.34$^{+0.23}_{-0.27}$ &  1,2,3 \\
    Year 4 & 1992 Jan-1992 Oct & 42.97$\pm$0.17 & 11.00$^{+2.00}_{-1.90}$ & 5986$\pm$95 & 5691$\pm$164 & 3109$\pm$53 & 2264$\pm$88 & 0.014 & 1.10$^{+0.22}_{-0.21}$ &  1,2,3 \\
    Year 5 & 1992 Nov-1993 Sep & 43.24$\pm$0.07 & 13.00$^{+1.40}_{-1.60}$ & 5930$\pm$42 & - & 2486$\pm$13 & 1909$\pm$129 & 0.027 & 0.93$^{+0.16}_{-0.17}$ &  1,2,3 \\
    Year 6 & 1993 Nov-1994 Oct & 43.25$\pm$0.09 & 13.40$^{+4.30}_{-3.80}$ & 7378$\pm$39 & 7202$\pm$392 & 2877$\pm$17 & 2895$\pm$114 & 0.028 & 2.19$^{+0.72}_{-0.65}$ &  1,2,3 \\
    Year 7 & 1994 Nov-1995 Oct & 43.40$\pm$0.06 & 21.70$^{+2.60}_{-2.60}$ & 6946$\pm$79 & 6142$\pm$289 & 2432$\pm$13 & 2247$\pm$134 & 0.039 & 2.14$^{+0.36}_{-0.36}$ & 1,2,3 \\
    Year 8 & 1995 Nov-1996 Oct & 43.28$\pm$0.12 & 16.40$^{+1.10}_{-1.20}$ & 6623$\pm$93 & 5706$\pm$357 & 2276$\pm$15 & 2026$\pm$68 & 0.030 & 1.31$^{+0.12}_{-0.13}$ &  1,2,3 \\
    Year 9 & 1996 Dec-1997 Oct & 43.08$\pm$0.10 & 17.50$^{+1.60}_{-2.00}$ & 6298$\pm$65 & 5541$\pm$354 & 2178$\pm$12 & 1923$\pm$62 & 0.019 & 1.26$^{+0.14}_{-0.17}$ & 1,2,3 \\
    Year 10 & 1997 Nov-1998 Sep & 43.43$\pm$0.08 & 26.50$^{+2.20}_{-4.30}$ & 6177$\pm$36 & 4596$\pm$505 & 2035$\pm$11 & 1732$\pm$76 & 0.041 & 1.55$^{+0.19}_{-0.29}$ & 1,2,3 \\
    Year 11 & 1998 Nov-1999 Oct & 43.34$\pm$0.11 & 24.80$^{+3.00}_{-3.20}$ & 6247$\pm$57 & 6377$\pm$147 & 2021$\pm$18 & 1980$\pm$30 & 0.034 & 1.90$^{+0.24}_{-0.25}$ & 1,2,3 \\
    Year 12 & 1999 Dec-2000 Sep & 42.89$\pm$0.21 & 6.50$^{+3.70}_{-5.70}$ & 6240$\pm$77 & 5957$\pm$224 & 2010$\pm$30 & 1969$\pm$48 & 0.012 & 0.49$^{+0.28}_{-0.43}$ &  1,2,3 \\
    Year 13 & 2000 Nov-2001 Dec & 42.87$\pm$0.16 & 14.30$^{+7.30}_{-5.90}$ & 6478$\pm$108 & 6247$\pm$343 & 3111$\pm$131 & 2173$\pm$89 & 0.011 & 1.32$^{+0.68}_{-0.55}$ & 1,2,3 \\
    Year 17 & 2005 Mar-2005 Apr & 42.49$\pm$0.24 & 6.30$^{+2.30}_{-2.60}$ & 6396$\pm$167 & - & 3210$\pm$642 & 2939$\pm$373 & 0.005 & 1.06$^{+0.47}_{-0.51}$ & 3,4 \\
    Year 19 & 2007 Mar-2007 Jul & 42.64$\pm$0.15 & 12.40$^{+3.85}_{-2.74}$ & 11481$\pm$574 & 4849$\pm$112 & 4212$\pm$211 & 1822$\pm$35 & 0.007 & 0.80$^{+0.25}_{-0.18}$ & 3,5 \\
    Year 20\textsuperscript{b} & 2008 Feb-2008 Jun & 42.59$\pm$0.04 & 4.17$^{+1.33}_{-0.90}$ & 12771$\pm$71 & 11177$\pm$2266 & 4266$\pm$65 & 4270$\pm$292 & 0.006 & 1.48$^{+0.51}_{-0.38}$ & 3,6 \\
    Year 23\textsuperscript{c} & 2012 Jan-2012 Apr & 43.38$\pm$0.05 & 2.83$^{+0.96}_{-0.88}$ & 1094$\pm$10 & 7038$\pm$122 & 3056$\pm$4 & 2772$\pm$33 & 0.017 & 0.42$^{+0.14}_{-0.13}$ & 7 \\
    Year 25 & 2014 Jan-2014 Jul & 43.04$\pm$0.07 & 4.17$^{+0.36}_{-0.36}$ & 9496$\pm$418 & 10161$\pm$587 & 3691$\pm$162 & 4278$\pm$671 & 0.037 &  1.49$^{+0.48}_{-0.48}$ & 8 \\
    Year 26 & 2015 Jan-2015 Aug & 43.14$\pm$0.10 & 7.20$^{+0.35}_{-1.33}$ & 11623$\pm$352 & 10241$\pm$515 & 4307$\pm$150 & 4377$\pm$477 & 0.021 & 2.69$^{+0.60}_{-0.77}$ & 9,10 \\
    Year 29 & 2018 Mar-2018 Jun & 43.22$\pm$0.07 & 7.01$^{+3.36}_{-2.33}$ & 12221$\pm$535 & 9724$\pm$599 & 3881$\pm$111 & 3989$\pm$429 & 0.025 & 2.18$^{+1.14}_{-0.86}$ & 10 \\
    Year 30 & 2018 Nov-2019 Jun & 43.30$\pm$0.08 & 8.89$^{+1.05}_{-2.03}$ & 10493$\pm$258 & 9053$\pm$710 & 3717$\pm$94 & 3732$\pm$300 & 0.031 & 2.42$^{+0.48}_{-0.68}$ & 10 \\
    Year 31 & 2020 Jan-2020 Jun & 43.40$\pm$0.04 & 10.03$^{+3.27}_{-3.28}$ & 9657$\pm$617 & 9199$\pm$863 & 3758$\pm$14 & 3209$\pm$600 & 0.039 & 2.02$^{+1.00}_{-1.00}$ & 10 \\
    Year 32 & 2020 Dec-2021 Aug & 43.34$\pm$0.08 & 9.02$^{+2.48}_{-1.90}$ & 9578$\pm$159 & 8470$\pm$423 & 3574$\pm$28 & 3349$\pm$436 & 0.034 & 1.98$^{+0.75}_{-0.66}$ & 10 \\
\enddata
\tablecomments{
$^{a}$ $L_{5100}$ and its uncertainties are calculated using the AGN continuum flux at 5100\AA~ ($F_{5100}$) and its standard deviation in Table 6 of \citet{Lu_2022}, with the cosmological constants referred in introduction. Due to the difference of $\text{H}_{0}$, the calculated $L_{5100}$ is slightly less than which in \citet{Lu_2016}. \\
$^{b}$ \cite{Pancoast_2014b} modeled the data and provided the model-dependent H$\beta$ lag of $3.22^{+0.66}_{-0.54}$ days, which is consistent (within the uncertainties) with $4.17^{+0.90}_{-1.33}$ days determined by the cross correlation analysis in \cite{Bentz_2010}. We use the H$\beta$ lag of \cite{Bentz_2010} for consistency.\\
$^{c}$ The FWHM measured from mean spectrum in Year 23 is much lower than which from the rms spectrum and other measurements, probably indicating that it may be unreliable. This FWHM$_\text{mean}$ is not adopted in the following analysis. \\
{References.} (1) \citet{Kilerci_2015}, (2) \citet{Colin_2006}, (3) \citet{Bentz_2013}, (4) \citet{Bentz_2007}, (5) \citet{Denny_2010}, (6) \citet{Bentz_2009}, (7) \citet{De_rosa_2018}, (8) \citet{Pei_2017}, (9) \citet{Lu_2016}, (10) \citet{Lu_2022}.
}
\end{deluxetable}

\begin{deluxetable}{ccccccccccc}
  \renewcommand\arraystretch{1.382}
  \tablecolumns{11}
  \tabletypesize{\scriptsize}
  \setlength{\tabcolsep}{4pt}
  \tablewidth{4pt}
  \tablecaption{Properties of NGC 4151 with measured $\tau_\text{cent}$ and velocities of the broad H${\beta}$ line for calibration.\label{tab:table2}}
  \tablehead{
\colhead{Data set}    &
\colhead{Duration}    &
\colhead{log$L_{\text{5100}}$}    &
\colhead{$\tau_{\text{H}\beta}$}    &
\colhead{FWHM$_\text{mean}$}    &
\colhead{FWHM$_\text{rms}$}    &
\colhead{$\sigma_\text{mean}$}    &
\colhead{$\sigma_\text{rms}$}    &
\colhead{$\lambda_\text{Edd}$}    &
\colhead{VP$_{\sigma,\text{r}}$}  &
\colhead{Ref}\\
\colhead{}    &
\colhead{}    &
\colhead{log(erg s$^{-1}$)}    &
\colhead{(days)}    &
\colhead{(km~s$^{-1}$)}    &
\colhead{(km~s$^{-1}$)}    &
\colhead{(km~s$^{-1}$)}    &
\colhead{(km~s$^{-1}$)}    &
\colhead{}    &
\colhead{($10^{7}M_{\odot}$)}    & 
\colhead{}  \\ 
\colhead{(1)}    &
\colhead{(2)}    &
\colhead{(3)}    &
\colhead{(4)}    &
\colhead{(5)}    &
\colhead{(6)}    &
\colhead{(7)}    &
\colhead{(8)}    &
\colhead{(9)}    &
\colhead{(10)}   &
\colhead{(11)}
}
\startdata
    Year 1 \textsuperscript{a} & 1987 Dec-1988 Jul & 42.62$\pm$0.03 & 11.50$^{+3.70}_{-3.70}$ & - & 5713$\pm$675 & - & 1958$\pm$56 & 0.015 & 0.86$^{+0.28}_{-0.28}$ & 1,2 \\
    Year 7 & 1993 Nov-1994 Feb & 43.00$\pm$0.03 & 3.10$^{+1.30}_{-1.30}$ & - & 4248$\pm$516 & - & 1914$\pm$42 & 0.036 & 0.22$^{+0.09}_{-0.09}$ & 2,3 \\
    Year 8 & 1994 Dec-1995 Jun & 43.06$\pm$0.05 & 6.40$^{+3.90}_{-4.90}$ & 6630$\pm$29 & 4689$\pm$170 & 2315$\pm$3 & 2068$\pm$72 & 0.041 & 0.53$^{+0.33}_{-0.41}$ & 4\textsuperscript{b} \\
    Year 9 & 1995 Nov-1996 Jul & 43.10$\pm$0.05 & 11.90$^{+2.40}_{-1.80}$ & 6360$\pm$21 & 6723$\pm$184 & 2307$\pm$2 & 2300$\pm$31 & 0.045 & 1.23$^{+0.25}_{-0.19}$ & 4 \\
    Year 10 & 1996 Dec-1997 Jul & 43.03$\pm$0.03 & 16.60$^{+3.80}_{-3.80}$ & 5741$\pm$21 & 4122$\pm$1228 & 2093$\pm$2 & 2153$\pm$81 & 0.039 & 1.50$^{+0.36}_{-0.36}$ & 4 \\
    Year 11 & 1997 Nov-1998 Jul & 42.98$\pm$0.03 & 9.90$^{+3.60}_{-2.70}$ & 6069$\pm$22 & 4990$\pm$315 & 2062$\pm$3 & 1971$\pm$44 & 0.034 & 0.75$^{+0.28}_{-0.20}$ & 4 \\
    Year 12 & 1998 Dec-1999 Apr & 42.91$\pm$0.04 & 7.60$^{+3.00}_{-4.60}$ & 5714$\pm$21 & 4936$\pm$1001 & 2042$\pm$4 & 2181$\pm$67 & 0.030 & 0.71$^{+0.28}_{-0.43}$ & 4 \\
    Year 15\textsuperscript{c} & 2001 Dec-2002 Jul & 42.77$\pm$0.04 & 1.70$^{+8.70}_{-2.10}$ & 7352$\pm$34 & 6224$\pm$793 & 2042$\pm$4 & 2181$\pm$67 & 0.021 & 0.23$^{+1.16}_{-0.23}$ & 4 \\
    Year 16 & 2002 Dec-2003 Jul & 42.90$\pm$0.04 & 13.00$^{+6.60}_{-6.30}$ & 6224$\pm$24 & 5707$\pm$543 & 2187$\pm$4 & 2117$\pm$66 & 0.029 & 1.14$^{+0.58}_{-0.56}$ & 4 \\
    Year 17 & 2003 Nov-2004 Jul & 42.67$\pm$0.10 & 13.20$^{+3.30}_{-4.90}$ & 6118$\pm$42 & 4930$\pm$69 & 2409$\pm$10 & 2091$\pm$14 & 0.017 & 1.13$^{+0.28}_{-0.42}$ & 4 \\
    Year 19 & 2005 Mar-2005 Apr & 42.56$\pm$0.06 & 6.60$^{+1.10}_{-0.80}$ & - & 4711$\pm$750 & - & 2680$\pm$64 & 0.013 & 0.93$^{+0.16}_{-0.12}$ & 5 \\
    Year 26 & 2012 Jan-2012 May & 42.47$\pm$0.09 & 6.80$^{+0.50}_{-0.60}$ & 5174$\pm$32 & 4393$\pm$110 & 2078$\pm$2 & 1940$\pm$22 & 0.015 & 0.50$^{+0.04}_{-0.05}$ & 6 \\
    Year 32 & 2018 Nov-2019 Aug & 42.09$\pm$0.10 & 5.50$^{+1.10}_{-1.00}$ & 5366$\pm$18 & 3235$\pm$243 & 2030$\pm$2 & 1844$\pm$44 & 0.004 & 0.37$^{+0.08}_{-0.07}$ & 4 \\
    Year 33\textsuperscript{c} & 2019 Nov-2020 Mar & 42.39$\pm$0.03 & 7.00$^{+1.90}_{-1.90}$ & 5191$\pm$12 & 1838$\pm$227 & 1973$\pm$1 & 2067$\pm$61 & 0.009 & 0.58$^{+0.16}_{-0.16}$ & 4 \\
    Year 34 & 2020 Nov-2021 May & 42.61$\pm$0.05 & 6.80$^{+0.90}_{-1.20}$ & 4677$\pm$19 & 2108$\pm$302 & 1794$\pm$2 & 1740$\pm$66 & 0.010 & 0.40$^{+0.06}_{-0.08}$ & 4 \\
    Year 34 & 2020 Nov-2021 Jun & 42.66$\pm$0.09 & 4.95$^{+2.12}_{-1.20}$ & 4937$\pm$14 & 5950$\pm$636 & 2095$\pm$6 & 2623$\pm$54 & 0.016 & 0.67$^{+0.29}_{-0.16}$ & 7 \\
    Year 35 & 2021 Oct-2022 Aug & 42.60$\pm$0.20 & 12.30$^{+0.80}_{-1.00}$ & 5072$\pm$17 & 2501$\pm$53 & 1963$\pm$2 & 1767$\pm$10 & 0.014 & 0.75$^{+0.05}_{-0.06}$ & 4 \\
    Year 35 & 2021 Nov-2022 May & 42.60$\pm$0.18 & 9.77$^{+1.40}_{-3.55}$ & 5163$\pm$17 & 4665$\pm$56 & 2167$\pm$7 & 2092$\pm$42 & 0.014 & 0.84$^{+0.12}_{-0.31}$ & 7 \\
    Year 36 & 2022 Nov-2023 Jul & 42.85$\pm$0.05 & 7.32$^{+0.88}_{-1.12}$ & 5066$\pm$13 & 5369$\pm$331 & 2154$\pm$5 & 2596$\pm$42 & 0.026 & 0.96$^{+0.12}_{-0.15}$ & 7 \\
    Year 37 & 2023 Nov-2024 Jun & 42.94$\pm$0.04 & 3.25$^{+1.40}_{-0.72}$ & 5407$\pm$7 & 6146$\pm$536 & 2297$\pm$3 & 2765$\pm$43 & 0.031 & 0.49$^{+0.21}_{-0.11}$ & 7 \\
\enddata
\tablecomments{$^{a}$ The line widths of the broad line may be unreliable, due to a combination of narrow-line residuals in the rms spectra, uncertain narrow-line removal from the original spectra, and a variable line-spread function \citep{Peterson_2004}. \\
$^{b}$ Data in Ref. 4 comes respectively from the AGN Watch Project (Year 8 to Year 17) and the MAHA Program (Year 32 to Year 35). \\
$^{c}$ The $r_\text{max}$ from CCF is lower than 0.5, indicating the time lag is unreliable \citep{Chen_2023}. Data from Year 15 and Year 33 is not adopted in the following analysis.\\
{References.} (1) \citet{Metzroth_2006}, (2) \citet{Peterson_2004}, (3) \citet{Kaspi_1996},(4) \citet{Chen_2023}, (5) \citet{Bentz_2006}, (6) \citet{De_rosa_2018}, (7) \citet{Feng_2024}.
}
\end{deluxetable}

NGC 5548 has been reverberation mapped for 23 times since 1988, from the AGN watch (13/23) \citep{Peterson_2004}, the campaigns at Lijiang Observatory of Yunnan Observatories (5/23) \citep{Lu_2016,Lu_2022} and other studies (5/23) \citep{Bentz_2007,Bentz_2009,Denny_2010,Pei_2017,De_rosa_2018}. NGC 4151 has also been RM monitored for 20 times since 1987, from the AGN watch (8/20), MAHA program (4/20) \citep{Findlay_2016,Du_2018b, Chen_2023}, Lijiang Observatory campaigns (4/20) \citep{Li_2022,Feng_2024}, and other studies (4/20) \citep{Maoz_1991,Kaspi_1996,Bentz_2006,De_rosa_2018}.
For these two AGNs with  largest numbers of RM observations, the continuum luminosity at 5100 \AA\ ($L_{5100}$),  the centroid time lag of the broad \hb line ($\tau_{\rm \hb}$), the $\rm FWHM_{\hb}$ and $\sigma_{\rm \hb}$ from the mean and rms spectra for each RM are adopted from Table 4 in \citet{Lu_2016}, Table 6 in \citet{Lu_2022}, Table 3 in \citet{Chen_2023} and Table 4 in \citet{Feng_2024} (see Table \ref{tab:table1},  \ref{tab:table2} respectively for NGC 5548 and NGC 4151). 

For data from the Lijiang Observatory, $L_{5100}$ is derived by disentangling host and AGN components from the observed spectrum, including a stellar template with an age of 11 Gyr and metallicity of Z = 0.05 for host-galaxy starlight and AGN component (such as Gaussian functions for the broad and narrow emission lines, a power law for the continuum, an iron template from \cite{Boroson_1992} for Fe II emissions \citep{Lu_2022,Feng_2024}. 
For data from other campaigns, the flux contributions from the host galaxy were mainly estimated via two-dimensional image decomposition using GALFIT applied to Hubble Space Telescope (HST) Advanced Camera for Surveys (ACS) imaging data \citep{Bentz_2013}. The decomposition involved modeling the host galaxy components with S\'{e}rsic profiles and subtracting the nuclear point source using a point-spread function (PSF) model \citep{Kilerci_2015,Chen_2023}. 

For these two AGNs, $\rm log_{10}(\mbh/\msun)$ is  adopted as $7.59^{+0.24}_{-0.21}$  for NGC 5548 \citep{Pancoast_2014b} and $7.22^{+0.11}_{-0.10}$ for NGC 4151 \citep{Bentz_2022} from the BLRs dynamical model.
The mass accretion rate $\dot{M}$ can be derived from the optically thick, geometrically thin accretion disk model \citep{Shakura_1973}, which has been extensively applied for fitting AGN spectra \citep{Colin_2006,Davis_2011,Liu_2022}. Considering the radius distribution of the effective disk temperature is given by $T_{\rm eff} \propto R^{-3/4}$, $\dot{M}$ can be expressed as:
\begin{equation}
    \dot{M} = 0.82 \left( \frac{l_{44}}{\cos i} \right)^{3/2} m_7^{-1} \, M_\odot \, \text{yr}^{-1}
    \label{eq:Mdot}
\end{equation}
where $l_{44}=L_{5100}/(10^{44}~{\rm erg~ s}^{-1})$, $m_{7}=\mbh/(10^{7}\msun)$. An average value of $\text{cos}~i =0.75$ is adopted, which corresponds to the opening angle of the dusty torus \citep{Davis_2011,Du_2016}. 
The mass accretion rate calculated with Equation~\ref{eq:Mdot} for this two AGNs are both $\sim 0.01 \msun/{\rm yr}$, which means the \mbh can be considered as a constant in decades. 

\section{Data Analysis} \label{sec:3}

\subsection{The virial relation between \hb line width and the \hb lag} \label{subsec:3.1}

\begin{figure*}
    \centering
    \subfigure{\label{Fig:R11a}
    \includegraphics[angle=0,width=0.7\textwidth]{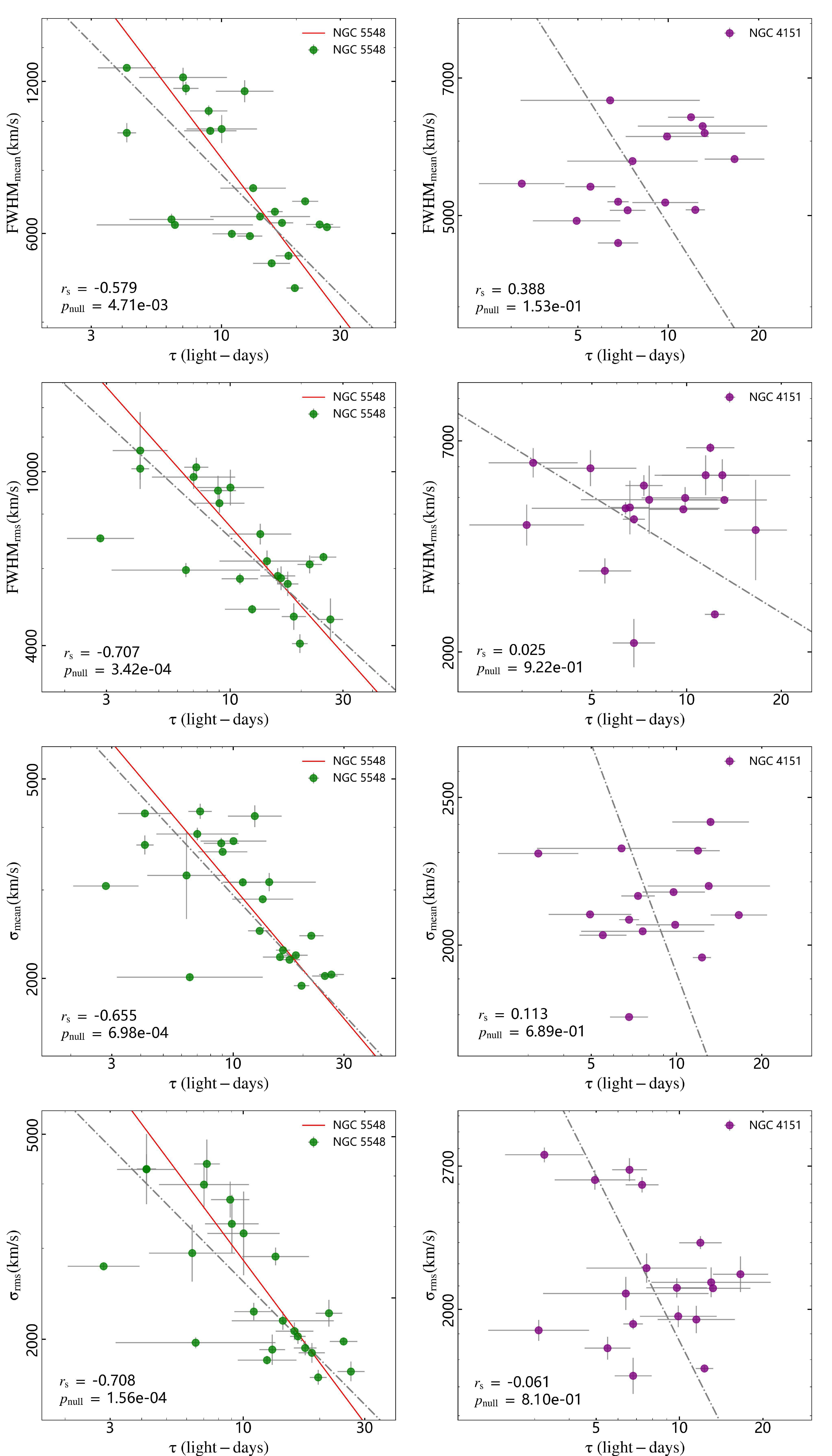}}
    \caption{Correlations between H${\beta}$ time lag and four kinds of velocity tracers for NGC 5548 (left panel) and NGC 4151 (right panel). Results of the Spearman correlation test are given at the bottom of each panel. The red solid line is the  best linear fitting for NGC 5548 and the results of the best fit are recorded in Table~\ref{tab:3}. The grey dotted-dashed lines are the best fits with a forced virial slope of $\alpha = -0.5$.}
    \label{Fig:1}
\end{figure*}

If $f$ is a constant,  there is a virial relation $\Delta V \propto R_{\rm BLR}^{-0.5} \propto \tau_{\rm \hb}^{-0.5} $ (see Equation \ref{eq:1}), where $\Delta V$ can be traced by FWHM or $\sigma_{\rm \hb}$ of  broad \hb emission line. It is possible that the line width obtained from the mean spectrum just represents the projected velocity of the BLR with the contributions of the broad-line flux, while the line width derived from the rms spectrum actually represents the projected velocity of the BLR associated with the broad-line variability (i.e., the variable region) \citep{Lu_2022}. Both the FWHM and $\sigma_\text{line}$ from the mean and rms spectrum (i.e., $\rm FWHM_{\rm mean},~ FWHM_{\rm rms},~ \sigma_{\rm mean},~ \sigma_{\rm rms}$) are taken into account for examining the virial relation between $\Delta V$ and $\tau(\rm \hb)$. 
Fig \ref{Fig:1} shows these virial relations for NGC 5548 (left) and NGC 4151 (right). From the top to the bottom panel, $\Delta V$ is traced by $\rm FWHM_{\rm mean},~ FWHM_{\rm rms},~ \sigma_{\rm mean},~ \sigma_{\rm rms}$, respectively.
The Spearman correlation coefficient $r_{\rm s}$ and the probability of null hypothesis $p_{\rm null}$ are also shown in panels. For relations with  $|r_\text{s}|>0.4$  and $p_{\rm null} < 0.05$, the best linear fitting through $lmfit$ is performed (red lines in Fig \ref{Fig:1}). The errors of parameters are modeled by the MCMC methods. The liner regression results are given in Table~\ref{tab:3}.
For NGC 5548, there are strong correlations between $\tau_{\rm \hb}$ and $\Delta V$ traced by $\Fm,~ \Fr, ~\sm,~ \sr$, with $r_s$ as $-0.58,~ -0.71,~ -0.66,~ -0.71$, respectively. The slopes are $-0.65,~-0.61,~-0.55,~-0.66$, which are deviated from the expected value of $-0.5$.  
For NGC 4151, these relations are not significant with $p_{\rm null}$ larger than 0.05.  

Therefore, assuming a constant $f$, the relations between $\tau_{\rm \hb}$ and four velocity tracers are all deviated from the virial  predictions for each AGN. $f$ is used to characterize the kinematics, geometry, and inclination of the BLR clouds.  The variable $f$  for each AGN  shows the variance  in properties of the BLR clouds in  different RM epochs.

RM studies have revealed a breathing mode of the broad Balmer lines such that when luminosity increases, the BLR size increases and the line width decreases \citep[e.g.,][]{WangS2020, Lu_2022, Chen_2023}. For a constant $f$, $ v^2 \thb \propto \mbh$,  the relation between $\thb$ and \lv is the same to the relation between the line width and \lv.
With  the Sloan Digital Sky Survey (SDSS) RM  project, \cite{WangS2020}  found that \hb displays the most consistent normal breathing expected from the virial relation,  \mgii and \ha on average show no breathing, and \civ (and similarly \ciii and \siiv) mostly shows anti-breathing. 
For the relation between $\thb$ and \lv, the Spearman correlation tests give a median strong correlation with $r_s=0.42$ and $p_{\rm null}=0.048$ for NGC 5548,  no significant relation with $r_s=0.124$ and $p_{\rm null}=0.62$ for NGC 4151. It shows a normal breathing in NGC 5548 and  complexity in breathing for NGC 4151.

From Equation \ref{eq:1}, the virial factor $f$ is calculated as $f=\mbh/VP$ with the data in Table \ref{tab:table1} and \ref{tab:table2}. Four kinds of $\Delta V$ tracers give four kinds of $f$ (i.e., $f_{\rm F,mean},~ f_{\rm F, rms},~ f_{\rm \sigma, mean},~ f_{\rm \sigma, rms}$). The errors of  $\log f$ can be calculated from the errors of $VP$ and $\mbh$:
\begin{equation}
    \delta \log f= \sqrt{(\delta \log VP)^2 + (\delta \log  \mbh)^2}
	\label{eq:error1}
\end{equation}
As $M_\text{BH}$ remains invariant over decades for each object, it can be considered as a constant and the errors can be neglected in the following analysis, that is, $\delta~\log~\mbh=0$ is adopted. The errors of $\log VP$ can be derived from  the errors of  $\Delta V$ and $\tau_{\rm \hb}$ as follows:
\begin{equation}
    \delta \log VP=\frac{1}{\ln(10)}  \sqrt{(\frac{\delta \tau_{\rm \hb}}{\tau_{\rm \hb}})^{2}+(2\frac{\delta \Delta V}{\Delta V})^{2}}
	\label{eq:error2}
\end{equation}

\begin{figure*}
    \centering
    \subfigure{\label{Fig:R21}
        \includegraphics[angle=0,width=0.8\textwidth]{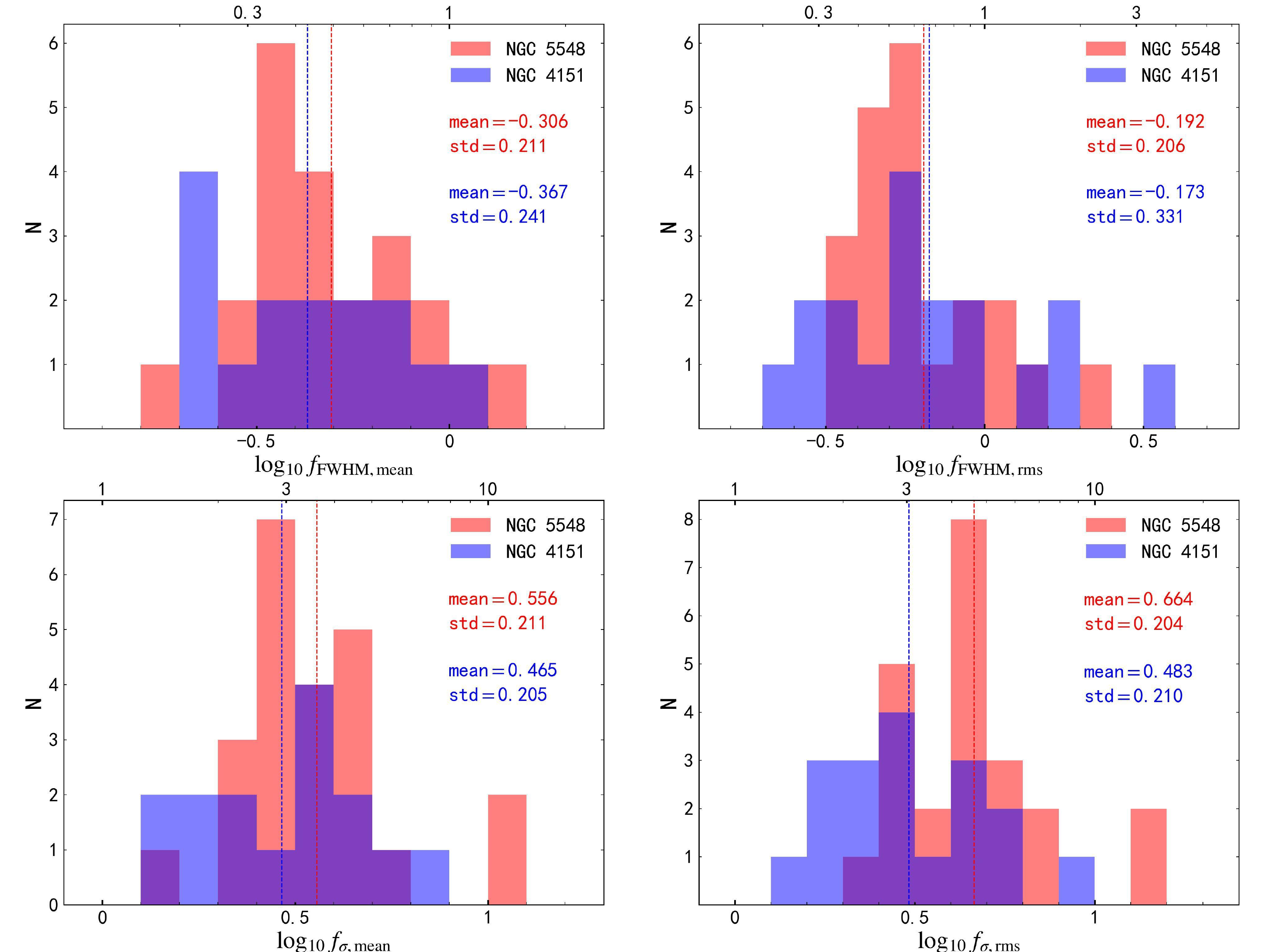}}
    \caption{The distributions of log$f$ calculated with four kinds of velocities. The red shaded bars are the distributions for NGC 5548 and blue for NGC 4151. The mean value and the standard deviation of log$f$ are shown in the panels. The dotted lines give the mean values for both targets.}
    \label{Fig:2}
\end{figure*}

\begin{deluxetable*}{ccccccccc}
\tablecolumns{9}
\renewcommand\arraystretch{1.382}
\tabletypesize{\scriptsize}
\setlength{\tabcolsep}{3pt}
\tablewidth{2pt}
\tablecaption{Linear regression results between the \hb broad line widths and time lags for NGC 5548 and NGC 4151.
\label{tab:3}}
\tablehead{
& \multicolumn{4}{c}{log$_{10} \tau$ (NGC 5548)} &\multicolumn{4}{c}{log$_{10} \tau$ (NGC 4151)} \\
    \cmidrule(lr){2-5} \cmidrule(lr){6-9} 
     & $\alpha$ & $\beta$ &$r_\text{s}$ & $p_\text{null}$ & $\alpha$ & $\beta$ &$r_\text{s}$ & $p_\text{null}$ 
}
\startdata
log$_{10}$F$_\text{m}$ & -0.65$\pm$0.04 & 4.58$\pm$0.05 & -0.58 & 0.01 & - & - & 0.46 & 0.07  \\
log$_{10}$F$_\text{r}$ & -0.61$\pm$0.05  & 4.49$\pm$0.06  & -0.71  & 0.00 & - & - & 0.03 & 0.92  \\
log$_{10}$$\sigma_{\text{m}}$ & -0.55$\pm$0.03 & 4.03$\pm$0.04 & -0.66 & 0.00 & - & - & 0.13 & 0.64  \\
log$_{10}$$\sigma_{\text{r}}$ & -0.66$\pm$0.07 & 4.12$\pm$0.08  & -0.71 & 0.00 & - & - & -0.04 & 0.89  \\
\enddata
\tablecomments{\footnotesize $\alpha$ and $\beta$ respectively represent the slope and the intercept of the regressions. The label 'm' and 'r' separately represent the velocities measured from the mean and rms spectrum. A liner fit will be given in case the absolute value of $r_\text{s}$ is greater than 0.4 and $p_\text{null}$ is less than 0.05. Symbols in the following Tables are the same as this.
}
\end{deluxetable*}

In Fig~\ref{Fig:2}, we show the distributions of four kinds of $f$, where the red is for NGC 5548 and the blue is for NGC 4151. The mean values of $\log f$ for $f_{\rm F,mean},~ f_{\rm F, rms},~ f_{\rm \sigma, mean},~ f_{\rm \sigma, rms}$ are  $-0.31,~ -0.19,~ 0.56,~ 0.66$ for NGC 5548, and  $-0.37,~ -0.17,~ 0.47,~ 0.48$ for NGC 4151. 
Four kinds of $f$ all have a range of about $1$ dex from the minimum to the maximum. While $\mbh$ of these two AGNs remain relative stable over four decades, the results above show that $f$ should not remain constant not only for sample of RM AGNs, but also for different RM epochs of an individual AGN.

\subsection{The relations between $f$ and the broad $\rm \hb$ line width \label{subsec:3.2}}

\begin{figure*}
    \centering
    \subfigure{\label{Fig:R31}
        \includegraphics[angle=0,width=0.7\textwidth]{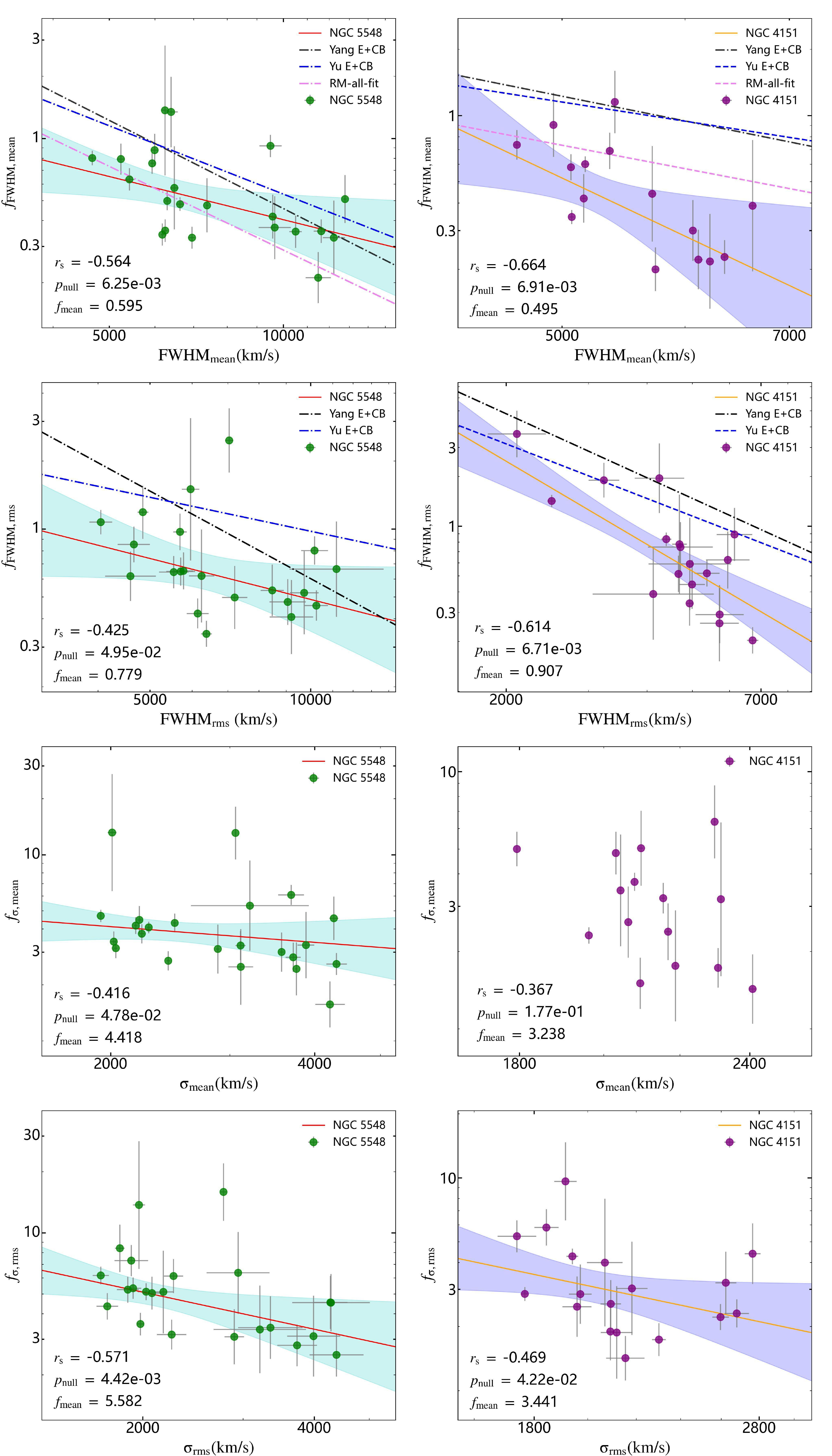}}
    \caption{Correlations between $f$ and the corresponding broad line widths for NGC 5548 and NGC 4151. The red and orange solid lines are the best linear fittings. The blue and purple shaded areas give the 95\% confidence intervals of the linear regressions. The coefficient of the Spearman correlation and the mean value of $f$ are exhibited in lower left of each panel. The black, blue and pink dotted-dashed lines are the relations respectively from the Yang E+CB RM subsample \citep{Yang_2024}, Yu E+CB RM subsample \citep{Yu_2020} and Liu SDSS-RM subsample \citep{Liu_2022}.}
    \label{Fig:3}
\end{figure*}

\begin{deluxetable*}{cccccccccccccc}
\tablecaption{Linear regression results between $f$ and the spectral properties for NGC 5548 and NGC 4151. \label{tab:table4}}
\tablehead{
& & \multicolumn{4}{c}{log$_{10}$$\Delta V$} &\multicolumn{4}{c}{log$_{10}$($L_\text{Bol}$/$L_\text{Edd}$)} &\multicolumn{4}{c}{log$_{10}$$D_{\text{H}\beta}$}\\
    \cmidrule(lr){3-6} \cmidrule(lr){7-10} \cmidrule(lr){11-14}
     & & $\alpha$ & $\beta$ &$r_\text{s}$ & $p_\text{null}$ & $\alpha$ & $\beta$ &$r_\text{s}$ & $p_\text{null}$ & $\alpha$ & $\beta$ &$r_\text{s}$ & $p_\text{null}$
}
\startdata
& log$_{10}$f$_\text{F,m}$ & -0.70$\pm$0.13 & 2.41$\pm$0.48 & -0.56 & 0.01 & - & - & -0.36 & 0.10 & -2.47$\pm$0.60 & 0.75$\pm$0.26 & -0.52 & 0.01 \\
NGC 5548 & log$_{10}$f$_\text{F,r}$ & -0.60$\pm$0.16  & 2.09$\pm$0.61  & -0.43  & 0.05 & -0.87$\pm$0.16 & -1.55$\pm$0.26 & -0.59 & 0.01 & - & - & -0.08 & 0.74\\
      & log$_{10}$f$_{\sigma,\text{m}}$ & -0.28$\pm$0.13 & 1.54$\pm$0.44 & -0.42 & 0.05 & - & - & -0.23 & 0.30 & - & - & 0.01 & 0.97 \\
& log$_{10}$f$_{\sigma,\text{r}}$ & -0.61$\pm$0.17 & 2.72$\pm$0.56  & -0.57 & 0.00 & -0.50$\pm$0.13 & -0.12$\pm$0.21 & -0.55 & 0.01 & - & - & 0.31 & 0.17\\
\hline
& log$_{10}$f$_\text{F,m}$ & -3.31$\pm$0.59 & 11.95$\pm$2.21 & -0.66 & 0.01 & - & - & -0.38 & 0.19 & -4.81$\pm$1.20 & 1.62$\pm$0.49 & -0.56 & 0.03\\
NGC 4151 & log$_{10}$f$_\text{F,r}$ & -1.67$\pm$0.16  & 5.90$\pm$0.56  & -0.61  & 0.01 & -1.02$\pm$0.15 & -2.34$\pm$0.32 & -0.51 & 0.04 & -2.74$\pm$0.38 & 0.71$\pm$0.13 & -0.75 & 0.00 \\
& log$_{10}$f$_{\sigma,\text{m}}$& - & - & -0.37 & 0.18 & - & - & -0.21 & 0.46 & - & - & -0.38 & 0.17\\
& log$_{10}$f$_{\sigma,\text{r}}$ & -1.16$\pm$0.28 & 4.31$\pm$0.94  & -0.47 & 0.04 & - & - & -0.40 & 0.11 & - & - & -0.39 & 0.11\\
\enddata
\tablecomments{\footnotesize $\alpha$ and $\beta$ respectively represent the slope and the intercept of the regressions. }
\end{deluxetable*}

Fig~\ref{Fig:3} shows the relations between four kinds of $f$  ( i.e., $f_{\rm F,mean},~ f_{\rm F, rms},~ f_{\rm \sigma, mean},~ f_{\rm \sigma, rms}$) and the corresponding broad \hb line width. 
There are negative strong correlations between four kinds of $f$ and corresponding $\Delta V$ with $p_{\rm null } <0.05$ and with $r_s$ as $-0.56,~ -0.43,~ -0.42,~ -0.57$ for NGC 5548. For NGC 4151, there are negative strong correlations between $f_{\rm F,mean},~ f_{\rm F, rms},~ f_{\rm \sigma, rms}$ and corresponding $\Delta V$ with $p_{\rm null } <0.05$ and  with $r_s$ as  $-0.66,~ -0.61,~ -0.47$ , except for the relation between $ f_{\rm \sigma, mean}$ and $\sm$.

In the upper two panels of Fig~\ref{Fig:3},  for the relation of $\ffm \propto \rm FWHM_{mean}^\alpha$, the best linear fittings give $\alpha=-0.7\pm0.13$ for NGC 5548 (red line, left panel), and  $\alpha=-3.31\pm0.59$ for NGC 4151 (orange line, right panel). In Fig~\ref{Fig:3}, the blue dotted-dashed line is for $\alpha=-1.10 \pm 0.4$ found by \cite{Yu_2020} for a sample of 17 Elliptical and Classical bulges (E+CB) RM AGNs. The black dotted-dashed line is for $\alpha=-1.42 \pm 0.26$ found by \cite{Yang_2024} for 40 E+CB RM AGNs with $f$ calibrated through $\mbh-L_{\rm bulge}$ relation. The pink dotted-dashed line is for $\alpha=-1.35 \pm 0.08$ found by \cite{Liu_2022} for a compiled sample of 120 RM AGNs.
When the velocity is traced by $\Fm$, the slope of $-0.7\pm0.13$  for NGC 5548 is larger than other three results, but the trend is similar, i.e., larger $f$ with smaller $\Fm$. Taking errors into account, the results from NGC 5548 are consistent with the result found by \cite{Yu_2020} for the 17 E+CB RM sample. However, for NGC 4151, the slope of $-3.31\pm0.59$ becomes much steeper than the other three results, with a strong correlation with $r_{\rm s}=-0.66$. When velocity is traced by $\Fr$, the results are similar to which by $\Fm$ for both objects. This indicates that $f$ could not be treated as a constant when calculating virial \mbh with $\rm FWHM_{\hb}$ for both NGC 5548 and NGC 4151. The linear regression results and the Spearman correlation coefficients are listed  in Table~\ref{tab:table4}. 

In the bottom two panels in Fig \ref{Fig:3}, $f_{\sigma,{\rm rms}}$ is moderately relevant to $\sigma_{\rm rms}$  with $r_{\rm s}$ as $-0.57$ for NGC 5548 and $-0.47$ for NGC 4151. When it comes to $f_{\sigma,\rm mean}$, the relations become much weaker (two panels in the penultimate row). For NGC 5548, $p_{\rm null}$ is up to 0.05 and the slope $\alpha$ is $-0.28\pm 0.13$ close to 0, and for NGC 4151 $p_{\rm null}$ is up to 0.18, both indicating there be weaker or no relations between the $f_{\sigma,\text{mean}}$ and $\sigma_{\rm mean}$. In Fig \ref{Fig:1}, for NGC 5548, the $\sm-\thb$ slope is closest to $-0.5$. It seems that, for a single AGN, $f_{\sigma,{\rm rms}}$ could not be considered as a constant, while $f_{\sigma,{\rm mean}}$ is more likely to be invariable. 
Our results  suggest that the variable $f$ should also be considered when calculating virial \mbh with FWHM or $f_{\sigma,\text{rms}}$ for an individual AGN with  different RM epochs.

\subsection{The relation between $f$ and the Eddington ratio\label{subsec:3.3}}

\begin{figure*}
    \centering
    \subfigure{\label{Fig:Ra212}
    \includegraphics[angle=0,width=0.7\textwidth]{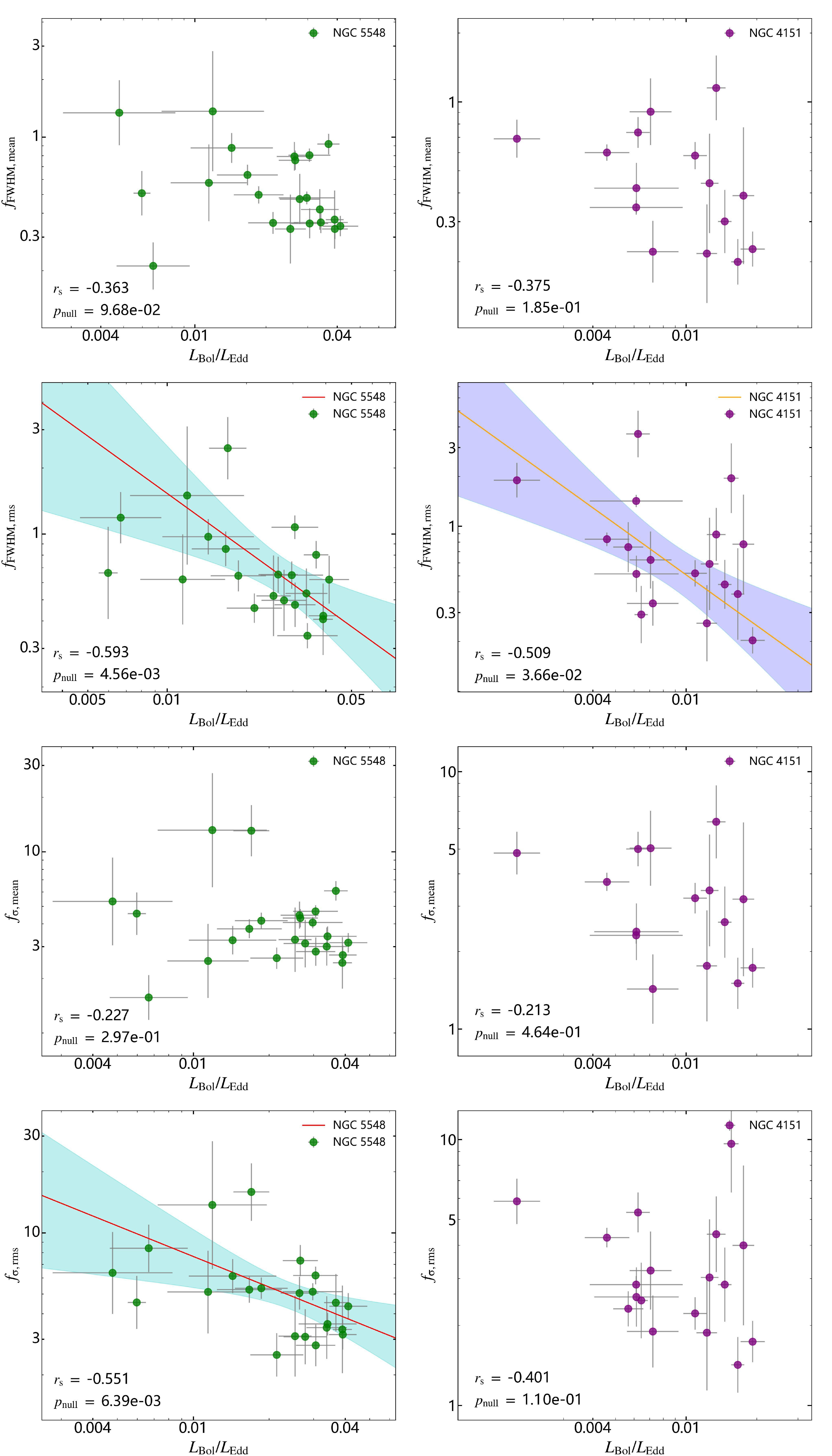}}
    \caption{Corelations between $f$ and the Eddington ratio for NGC 5548 and NGC 4151. Symbols, lines and the shaded areas are the same as which in Fig~\ref{Fig:3}.}
    \label{Fig:4}
\end{figure*}

The Eddington ratio $\lambda_{\rm Edd}$ ($\leddR$) is an important parameter describing the SMBH accretion process, where $L_{\rm Bol}$ and $L_{\rm Edd}$ are the bolometric luminosity and the Eddington luminosity, respectively. It depends on the estimations of \mbh and $L_{\rm Bol}$, where $L_{\rm Bol}= k_{5100} \times L_{5100}$. The bolometric correction coefficient $k_{5100}$ was suggested dependent on the luminosity or the Eddington ratio, here $k_{5100}=9$ is adopted \citep{Marconi_2004,Jin_2012,Wang_2019, Yu_2019}.

Fig \ref{Fig:4}  shows the relations between four kinds of $f$ and $\lambda_{\rm Edd}$ for these two AGNs. 
For NGC 5548, there is a strong anti-correlation between $f_{\rm F,rms}$ and $\leddR$ with $r_{\rm s}=-0.59,~ p_{\rm null}=4.56 \times 10^{-3}$. The best linear fitting gives $\ffr \propto (\leddR)^{-0.87\pm0.16}$. For NGC 4151, this correlation is still strong with $r_{\rm s}=-0.51$, $p_{\rm null}=3.66\times 10^{-2}$, and the best linear fitting gives $\ffr \propto (\leddR)^{-1.02\pm0.15}$. The relation indicates smaller $\ffr$ with larger $\leddR$. For an individual AGN, this relation also suggests smaller $\ffr$ with larger luminosity $L_{5100}$.
For a sample of 27 RM AGNs with \mbh from BLRs dynamical model \citep{Pancoast_2014b,Grier_2017a,Williams_2020,Bentz_2022}, \cite{Villafana_2023} found $\log f_{\rm F,rms}=(0.13^{+0.13}_{-0.14}) \log (\leddR)+0.32^{+0.19}_{-0.20}$, which is not consistent with our results of anti-correlation for an individual AGN. The slopes in linear regression by \cite{Villafana_2023} have large errors. Additional data are required to elucidate the distinction. 
For NGC 5548, we also find a correlation between $f_{\sigma,{\rm rms}}$ and $\leddR$ with $r_{\rm s} = -0.55,~ p_{\rm null}=0.01$.  
The correlations and the best fittings are shown in Table~\ref{tab:table4}. 

\subsection{The relation between $f$ and the broad \hb profile shape\label{subsec:3.4}}
\begin{figure*}
    \centering
    \subfigure{\label{Fig:Ra311}
        \includegraphics[angle=0,width=0.7\textwidth]{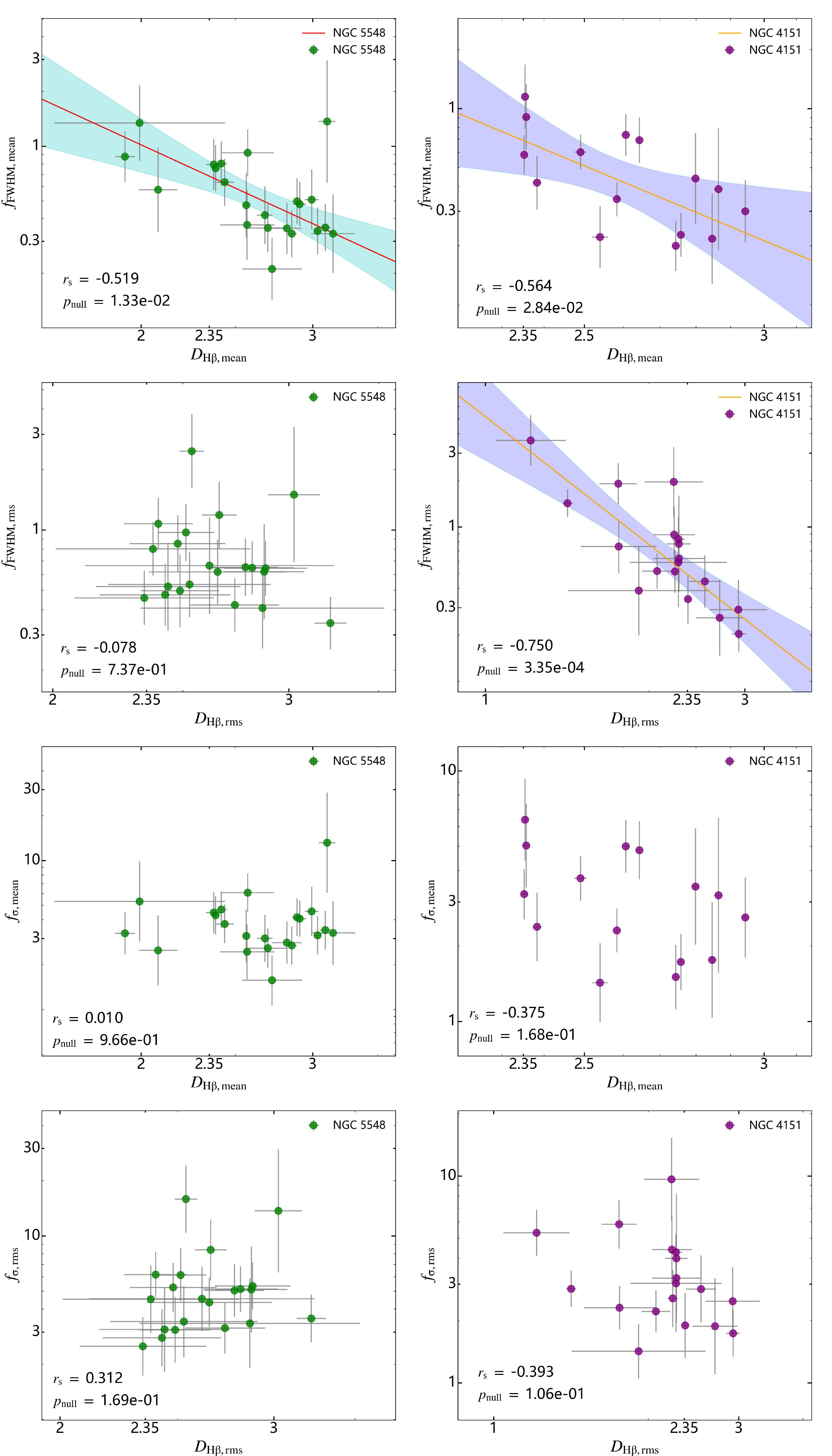}}
    \caption{Corelations between $f$ and the broad \hb line profile shape for NGC 5548 and NGC 4151. Symbols, lines and the shaded areas are the same as which in Fig~\ref{Fig:3}.}
    \label{Fig:5}
\end{figure*}

For the broad \hb line in the mean or rms spectra, the line profile shape is defined as $D_{\rm \hb} \equiv \rm FWHM_{\hb}/ \sigma_{\rm \hb}$. The value of $D_{\rm \hb}$ is 2.35, 3.46, 2.45, 2.83, and 0 for a Gaussian, a rectangular, a triangular, an edge-on rotating ring, and a Lorentzian profile, respectively. The value of $D_{\rm \hb}$ depends on the line profile and gives a relatively convenient parameter related to the dynamics of BLRs \citep{Colin_2006,Du_2016}.   
In Fig \ref{Fig:5}, we give the relation between four kinds of $f$ and the corresponding $D_{\rm \hb}$. The linear regression results are listed in Table~\ref{tab:table4}.  

From the mean spectra of NGC 5548, there is a moderate correlation between $f_{\rm F,mean}$ and $D_{\rm \hb,mean}$ with $r_{\rm s}=-0.52,~ p_{\rm null}=1.33\times 10^{-2}$, and the best fitting gives $\ffm \propto {D}_{\rm \hb, mean}^{-2.47\pm0.60}$. For NGC 4151, this correlation is stronger with $r_{\rm s}=-0.56,~ p_{\rm null}=2.84\times 10^{-2}$, and the best fitting gives $\ffm\propto {D}_{\rm \hb, mean}^{-4.84\pm1.20}$. 
For a sample of RM AGNs with BLRs dynamical model, \citet{Villafana_2023} found  the relation is  $\log f_{\rm F,mean}=(-0.28^{+0.93}_{-0.92})\log D_{\rm \hb} + 0.01^{+0.31}_{-0.32}$ with a large uncertainty. Our results are not consistent with theirs, although the trend is similar.
The slope difference is possibly due to a larger range of $D_{\rm \hb} $  in \citet{Villafana_2023} than ours.
For the rms spectra of NGC 4151, there is a strong anti-correlations between $f_{\rm FWHM,rms}$ and $D_{\rm \hb,rms}$  with $r_{\rm s}= -0.75$.
The relations between $f$ and the line profile shape show that, for individual objects, $D_{\rm \hb}$ are distinct in different RM epochs and $f$ is associated with the dynamics of BLRs.

$M_{\rm BH}$ for this two targets adopted here are derived from the dynamical modeling methods, which is dependent on the RM data sets. The  $M_{\rm BH}$  variation induces vertical offsets in the $f$ values. The correlation results between $f$ and observational parameters above remain unchangeable against the variation in $M_{\rm BH}$.

\section{Discussion}\label{sec:4}
\subsection{Cumulative fraction of $f$ and the BLR inclination \label{subsec:4.1}}

\begin{figure*}
    \centering
      \includegraphics[angle=0,width=0.4\textwidth]{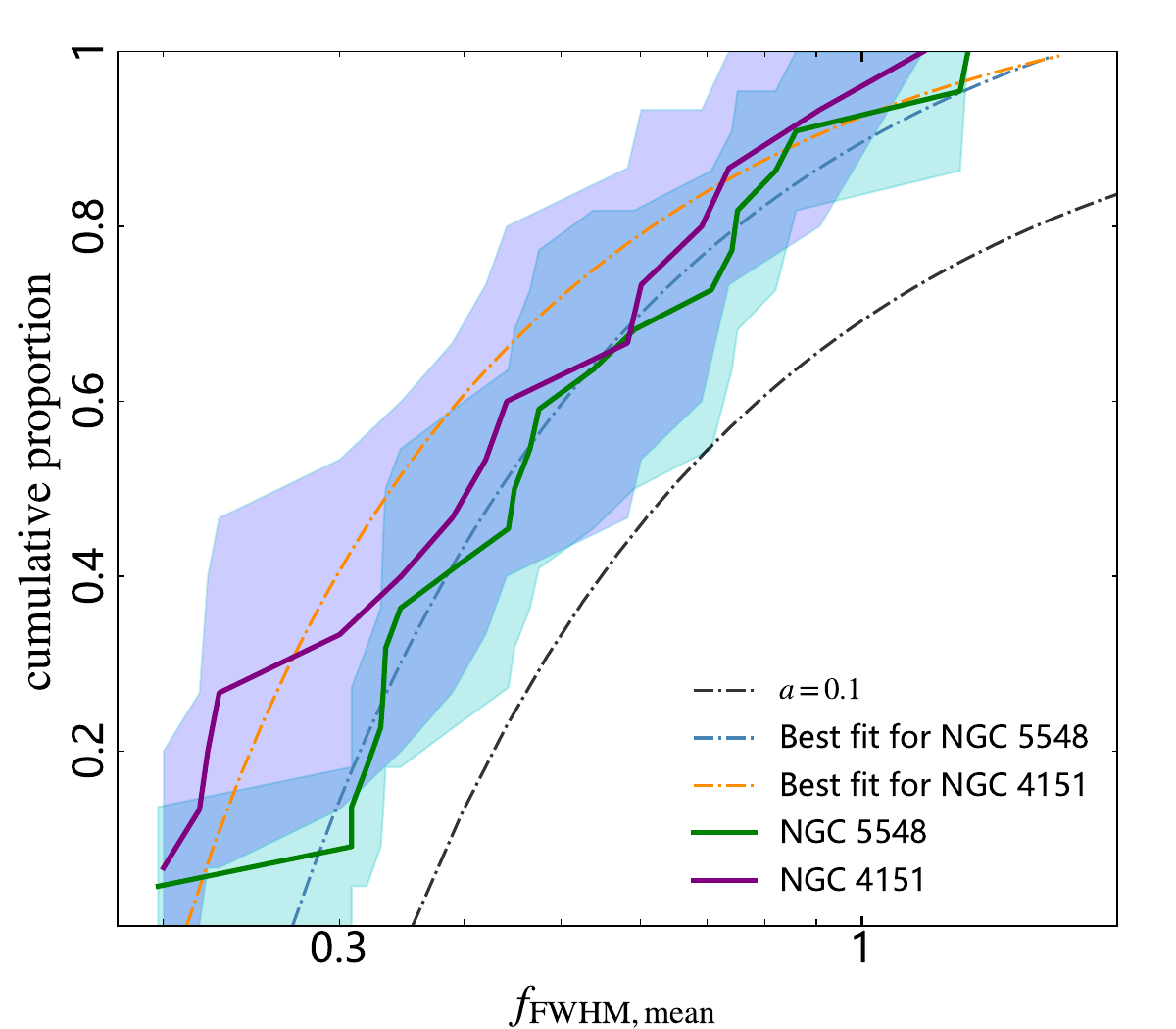}
      \includegraphics[angle=0,width=0.4\textwidth]{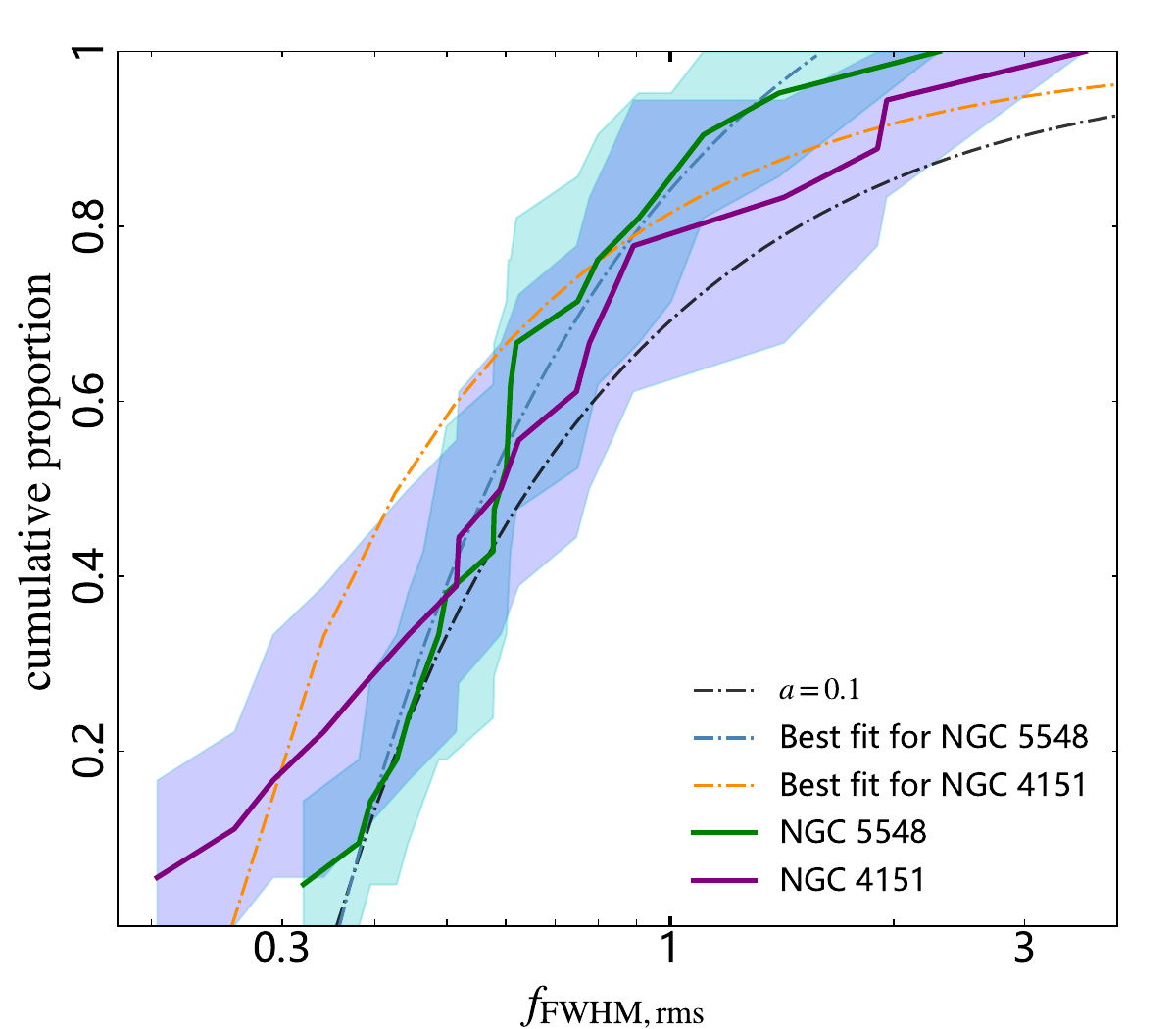}
      \includegraphics[angle=0,width=0.4\textwidth]{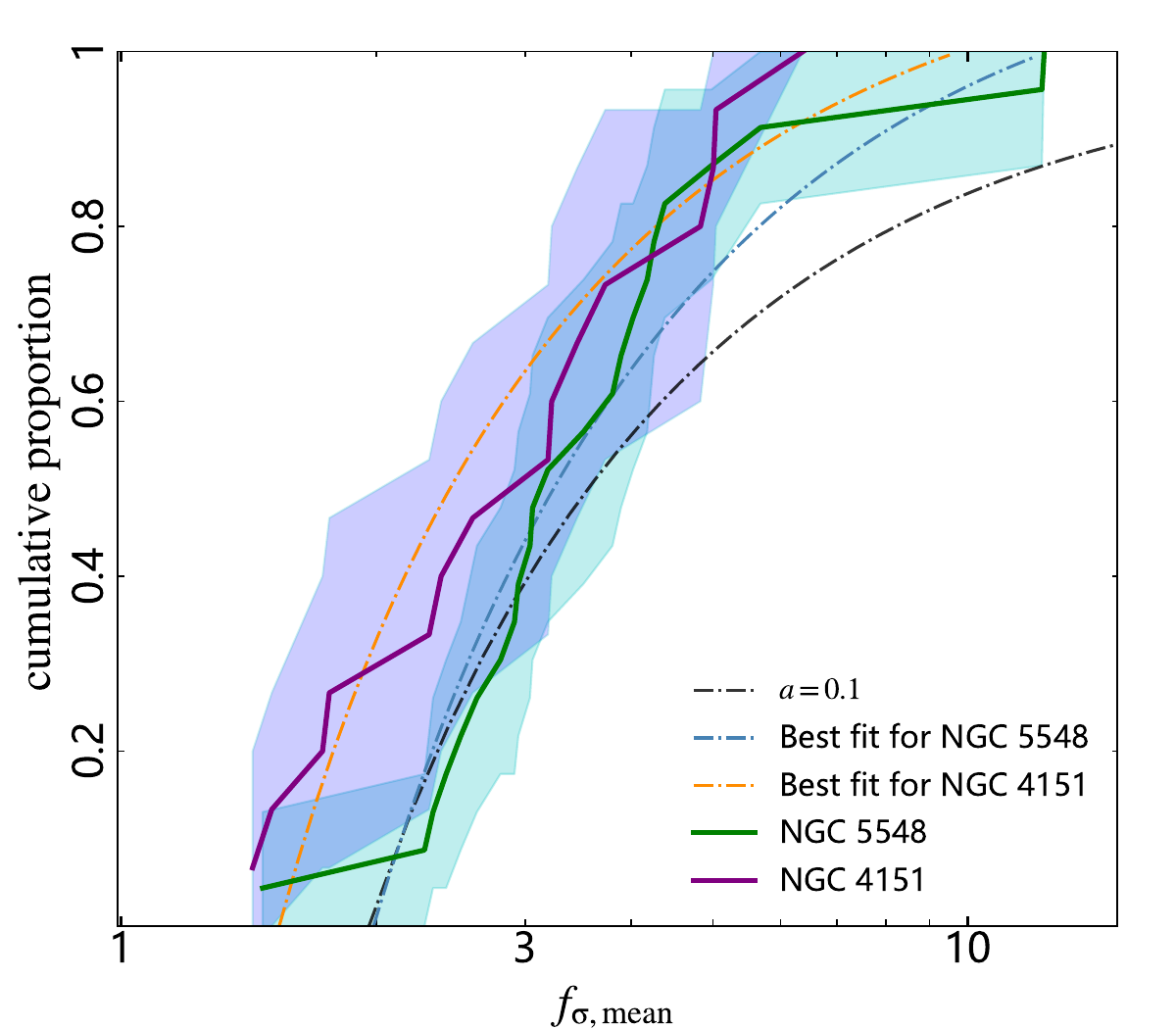}
       \includegraphics[angle=0,width=0.4\textwidth]{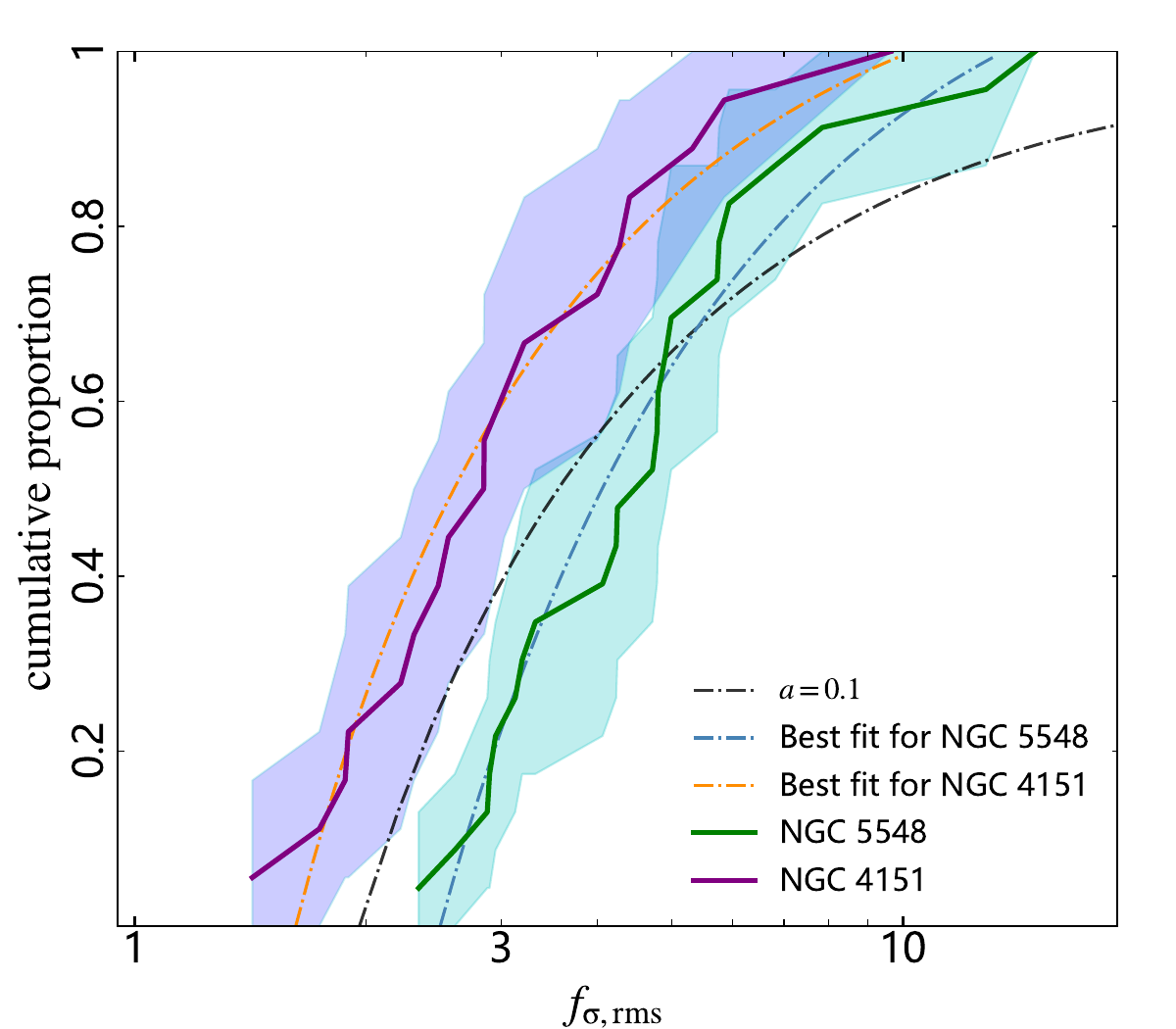}
    \caption{Cumulative fraction of four kinds of $f$ compared to the best fitting theoretical distributions. The green solid line is for NGC 5548 and purple line is for NGC 4151. The bootstrap method is used to obtain the confidence regions (90 per cent), which is shown as the green area (NGC 5548) and purple area (NGC 4151). The blue and yellow dotted-dashed lines are the best fit based on Equation~\ref{eq:f2}, which are calculated with $\sigma_\text{line}$. For FWHM, an offset factor has been divided to aid in comparison of the two distributions. For NGC 5548, the offset factor is 6.89, 6.97 respectively for the mean and rms spectrum, and 6.76, 4.15 for the mean and rms spectrum for NGC 4151, which are the squares of each mean $D_{\text{H}\beta}$. The black dotted-dashed line is the theoretical cumulative distribution constraining $a=0.1$, $\theta_\text{min}=0$ and $\theta_\text{max}=45~\text{deg}$, where the offset factor is 5.5 for both mean and rms spectrum. The fitting results are listed in Table~\ref{tab:5}.}
    \label{Fig:6}
\end{figure*}

\begin{deluxetable}{cccccc}
\tablecolumns{6}
\renewcommand\arraystretch{1.382}
 \tablewidth{4pt}
\tablecaption{Linear regression results for the theoretical cumulative distribution for NGC 5548 and NGC 4151.
\label{tab:5}}
\tablehead{
\colhead{}                 &
\colhead{}                  &
\colhead{$\theta_{\rm min}$}                   &
\colhead{$\theta_{\rm max}$}                       &
\colhead{$a$}                   &
\colhead{$\chi^{2}_{\rm min}$} 
}
\startdata
\hline
& $f_\text{F,m}$ & 14.18$\pm$4.57 & 45.27$\pm$2.11 & 0.185$\pm$0.102 & 0.64 \\
 & $f_\text{F,r}$ & 13.95$\pm$4.63  & 37.64$\pm$2.11  & 0.186$\pm$0.102  & 0.55 \\
NGC 5548 & $f_{\sigma,\text{m}}$ & 15.44$\pm$4.73 & 41.93$\pm$2.19 & 0.186$\pm$0.108 & 0.78 \\
&$f_{\sigma,\text{r}}$ & 12.71$\pm$3.98 & 37.63$\pm$1.72 & 0.166$\pm$0.091 & 0.66 \\
& mean & 14.07$\pm$4.48 & 40.62$\pm$2.03 & 0.181$\pm$0.101 & 0.66 \\
\hline
& $f_\text{F,m}$ & 16.60$\pm$3.45 & 56.11$\pm$2.16 & 0.111$\pm$0.078 & 0.91 \\
 & $f_\text{F,r}$ & 7.36$\pm$2.23 & 61.02$\pm$1.86 & 0.081$\pm$0.060 & 1.15 \\
NGC 4151 & $f_{\sigma,\text{m}}$& 16.92$\pm$4.41 & 53.28$\pm$3.46 & 0.165$\pm$0.106 & 0.93 \\
&$f_{\sigma,\text{r}}$ & 15.03$\pm$5.90 & 49.88$\pm$3.49 & 0.180$\pm$0.123 & 0.69 \\
& mean & 13.98$\pm$4.00 & 55.07$\pm$2.74 & 0.134$\pm$0.092 & 0.92 \\
\hline
\enddata
\end{deluxetable}

It was suggested that the virial factor $f$ has a significant relation with the inclination of the thick-disc of BLRs to the line of sight  \citep{Colin_2006,Yu_2019}. Assuming that BLRs consist of two dynamically distinct component, i.e., the disk and the wind, the observing velocity $v_\text{obs}$ could be expressed as follows \citep{Colin_2006}:
\begin{equation}
    v_\text{obs}=[k_{1}^{2}(a^{2}+\text{sin}^{2}\theta)V_\text{Kep}^{2}+k_{2}^{2}V_\text{out}^{2}\text{cos}^{2}\theta]^{1/2}
	\label{eq:v}
\end{equation}
where $a$ is the ratio of the scale height of the disk to the radius ($H/R$), or the ratio of the turbulent velocity to the local Keplerian velocity ($V_\text{turbulent}/V_\text{Kep}$), $V_\text{out}$ is the outflow velocity, which is assumed to be normal to the disk, $\theta$ is the inclination of the BLRs thick-disc to the line of sight, and $k_{1}$, $k_{2}$ are the contributions of the thick disc and of the wind, respectively. For a simple model of thick-disc BLRs, we neglect the contribution of outflow  in \hb profile, that is $k_{2}=0$  \citep[see Figure 11 in][]{Lu_2022}, the observing velocity could be written as $v_{\rm obs}=(a^{2}+{\rm sin}^{2} \theta)^{1/2}V_{\rm Kep}$. According to Equation~\ref{eq:1}, $f$ could be expressed as:

\begin{equation}
    f^\text{disk}=\frac{M_\text{BH}}{R_\text{BLR}v^{2}_\text{obs}/G}=\frac{R_\text{BLR}V^{2}_\text{Kep}/G}{R_\text{BLR}v^{2}_\text{obs}/G}=\frac{1}{a^{2}+\text{sin}^{2}\theta}
	\label{eq:f1}
\end{equation}
According to the unified scheme \citep{Antonucci_1985}, type I AGNs can be observed for an inclination less than $\theta_\text{max}$. The probability of seeing an object at an inclination angle $\theta$ per unit angle interval is $dN={\rm sin}\theta ~d \theta/({\rm cos} ~\theta_{\rm min} - {\rm cos}~ \theta_{\rm max})$, where $\theta_{\rm min}$ and $\theta_{\rm max}$ are the minimum and maximum angles that type I AGNs can be observed, associated with the maximum and the minimum $f$, respectively. 
With Equation~\ref{eq:f1}, the derivative of $f$ with respect to $\theta$ can be calculated, and the theoretical cumulative fraction of $f$ could be expressed as:
\begin{equation}
    N=\int_{f_\text{min}}^{f_\text{max}}\frac{1}{2(\text{cos} \ \theta_\text{min}-\text{cos} \  \theta_\text{max})}\frac{1}{f^{2}(1+a^{2}-f^{-1})^{1/2}}\text{d}f
	\label{eq:f2}
\end{equation} 

For samples of type I AGNs, cumulative fraction of four kinds of $f$ was used to investigate which kind of velocity tracer has dependence on the inclination \citep{Yu_2019}, where the parameters are often adopted as $\theta_\text{min}$ = 0, $\theta_\text{max}$ = 45 deg, a = 0.1 or 0.3. 
For an individual AGN, the variation of the distribution of BLRs clouds due to turbulent motion would possibly change the BLRs inclination during several decades.

Fig~\ref{Fig:6} shows the cumulative fraction of four kinds of $f$ for NGC 5548 (green solid line) and NGC 4151 (purple solid line). The bootstrap method is used to obtain the 90\% confidence regions and the standard deviation is calculated as the error estimation. 
There is an "offset factor" between FWHM-based $f$ and $\sigma$-based $f$, by which the FWHM-based $f$ has been divided to aid in comparison of the theoretical distribution of $f^\text{disk}$ \citep{Colin_2006}.
Here we adopt the square of the mean values of each $D_{\rm \hb}$ as the offset factor, that is 6.89, 6.97 respectively in the mean and rms spectra for NGC 5548, and 6.76, 4.15 for NGC 4151. 
With Equation~\ref{eq:f2}, from the cumulative fraction of four kinds of $f$,  $\theta_{\rm min}, \theta_{\rm max}$, and $a$ could be constrained. The best fit through $lmfit$ is reached when $\chi^{2}$ gets to the minimum (the dotted-dashed lines). 
For cumulative fractions of four velocity tracers, $\chi^{2}$ is around 1, which shows that the fittings are all acceptable.  $\chi^{2}$ about  $f_\text{F,r}$ is minimum of 0.55 for NGC 5548, and $\chi^{2}$ about  $f_{\sigma,\text{r}}$  is minimum of 0.69 for NGC 4151. The mean $\chi^{2}$  is 0.66 for NGC 5548 and 0.92 for NGC 4151.
The mean values in the best fitting are $\theta_{\rm min}=14.1 \rm ~deg$, $\theta_{\rm max}=40.6 \rm~ deg$, $a=0.18$ for NGC 5548, and $\theta_{\rm min}=14.0 \rm ~deg$, $\theta_{\rm max}=55.1 \rm~ deg$, $a=0.13$ for NGC 4151.
The fitting results are shown in Table~\ref{tab:5}. 

BLRs dynamical model provided the BLRs opening angle $\theta_{\rm o}$ and the disk inclination $\theta_i$ for these two AGNs \citep{Pancoast_2014b, Bentz_2022}, and the results are  $\theta_{\rm o}=27.4^{+10.6}_{-8.4}, \theta_i=38.8^{+12.1}_{-11.4}$ for NGC 5548 , and $\theta_{\rm o}=56.6^{+15.8}_{-14.3}, \theta_i=58.1^{+8.4}_{-9.6}$ for NGC 4151.
BLR in NGC 4151 has a large $\theta_{\rm max}$, which is $\sim 15~ {\rm deg}$ larger than NGC 5548, and this is consistent with the dynamical model results within errrors \citep{Bentz_2022}. NGC 5548 has larger $a$ than NGC 4151, indicating a higher proportion of turbulent motion in NGC 5548. Assuming that the BLR clouds are dominated by Kepler motion, and the clouds can reach everywhere of the region enclosed by the opening angle, 
the range of inclination $\theta$ is ($\theta_{\rm i} - \theta_{\rm o}$, $\theta_{\rm i} + \theta_{\rm o}$). The BLRs dynamical model gave the range of BLRs inclination $\theta$ is [11.4, 66.2] deg for NGC 5548, [1.5, 90] deg for NGC 4151. For NGC 4151, $\theta_{\rm min}$ can reach 1.5 deg and $\theta_{\max}$ can reach over 90 deg, which is different with our fitting results. For NGC 5548,  the lower limit of $\theta_{\rm min}$ is consistent with the BLRs dynamical  model results within errors, although there is  a difference for the upper limit. $\theta_{\max}$ for NGC 5548 is $\sim$15 deg less than NGC 4151.
Therefore, based on thick-disk BLRs model, the variation of  mean inclination leads to the variation of $f$. The inclination range is  $14.1-40.6$ and $14.0-55.1$ deg for NGC 5518 and NGC 4151, respectively.


\subsection{The dampling timescale of $f$ \label{subsec:4.2}}

\begin{figure*}
    \centering
        \includegraphics[width = 6in]{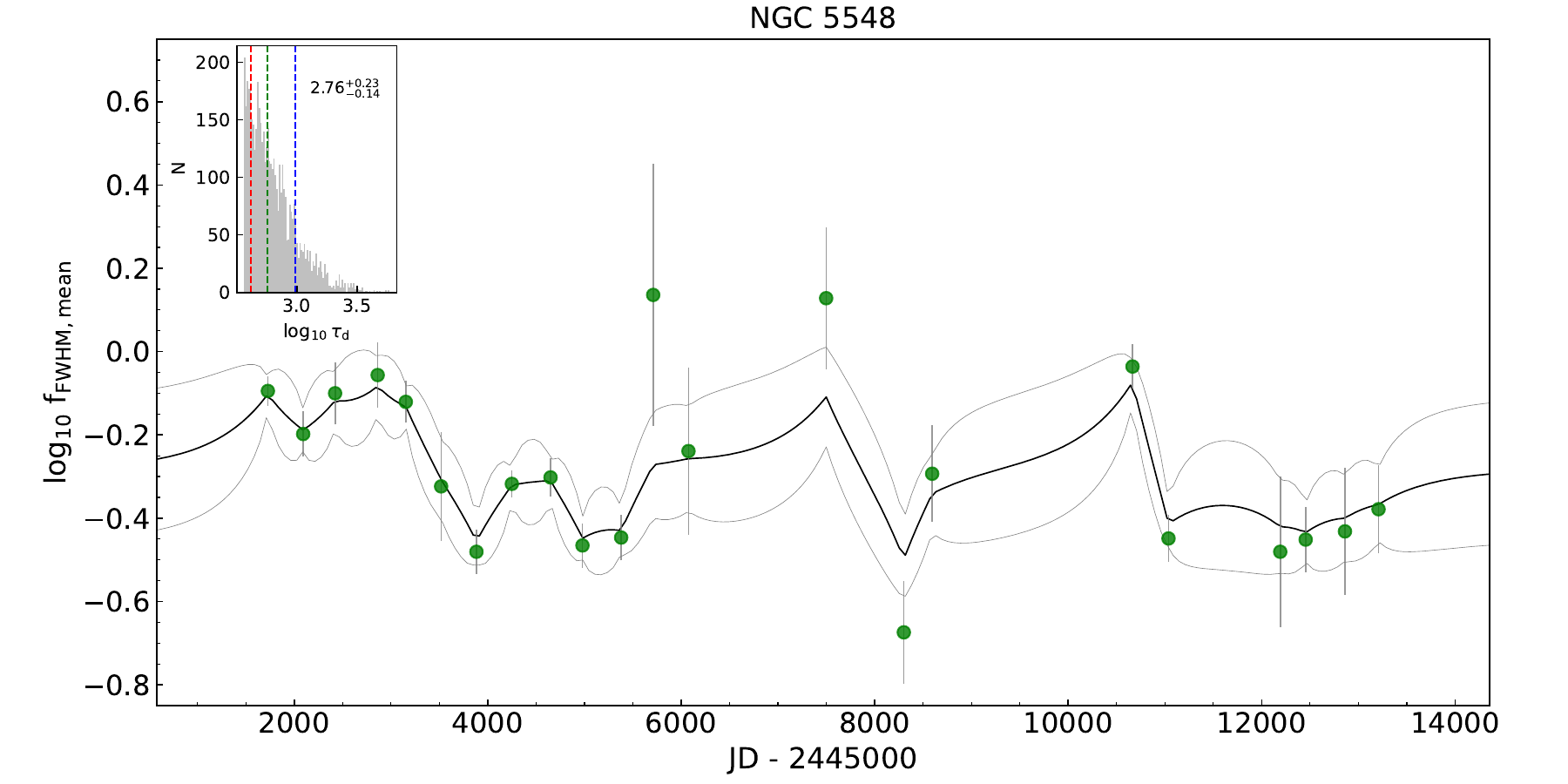}
        \includegraphics[width = 6in]{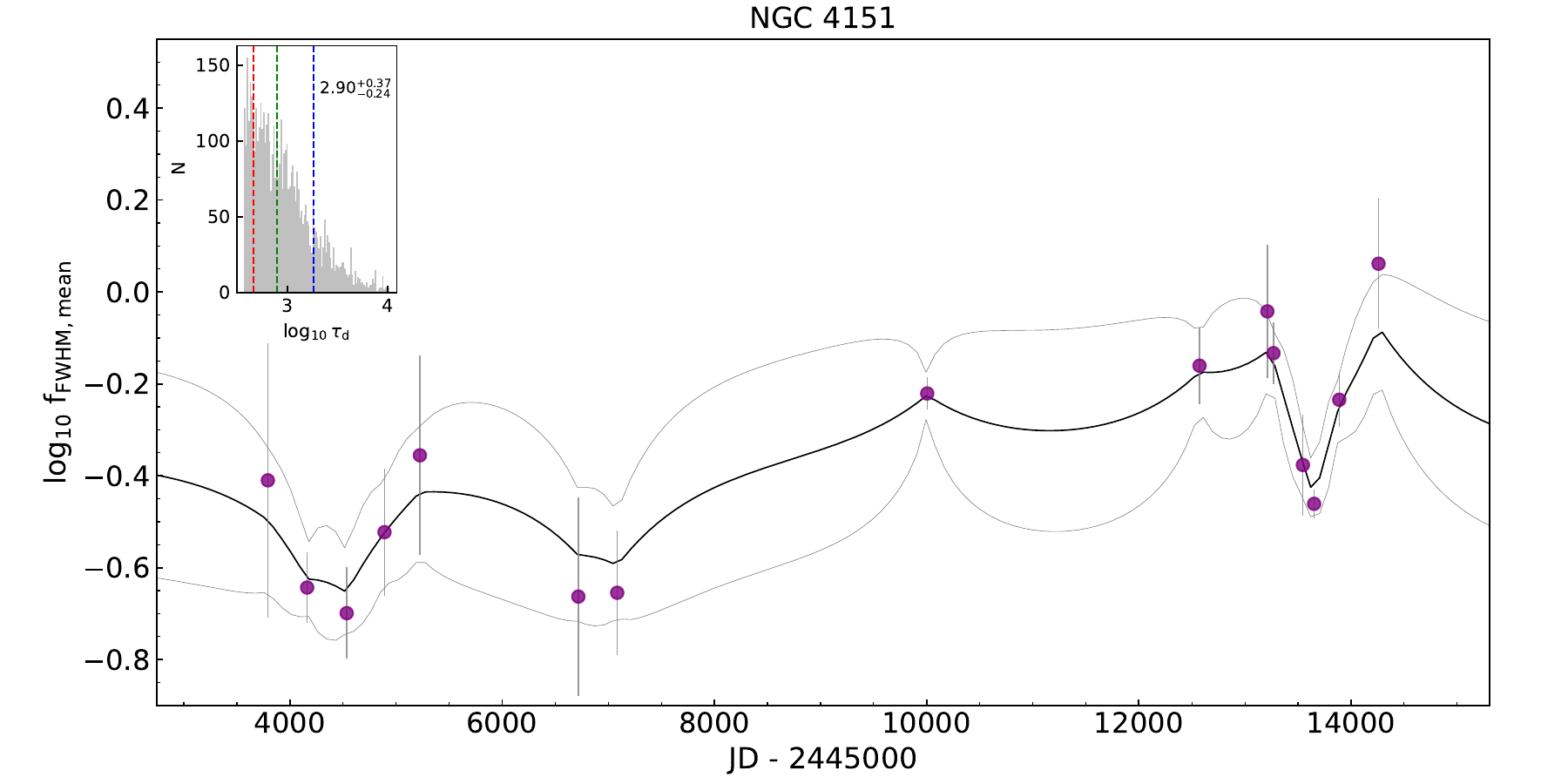}
    \caption{The DRW modelling results of log$\ffm$ for NGC 5548 and NGC 4151. The damping timescale $\tau_\text{D}$  is shown in the upper left panels of each picture.  The median value of $\text{log}_{10}\tau_\text{D}$ is depicted as the 50th percentile (green dashed line) in the panel, with the 16th (red dashed line) and
84th percentiles (blue dashed line) representing the lower and upper bounds of the error, respectively.}
    \label{Fig:7}
\end{figure*}

Damped random walk (DRW) is a stochastic process, defined by an exponential covariance function \citep[e.g.,][]{Zu_2011,Zu_2013,Lu_2019}:
\begin{equation}
	S(t_1, t_2) = \sigma_\text{D}^2 \exp\left( -\frac{|t_i - t_j|}{\tau_\text{D}} \right)
	\label{eq:DRW}
\end{equation}
where $\tau_\text{D}$ is the characteristic variability timescale, $t_i-t_j=\Delta t$ is the time sampling interval between the $i$th and $j$th observations, and $\sigma_\text{D}$ is the characteristic variability amplitude. It has been shown that AGN
variability in the optical band (e.g. $L_{5100}$) can be well described by the DRW model \citep{Kelly_2009,Zu_2013}. 
Fig \ref{Fig:7} shows the light curves of $\log \ffm$. The damping timescale is given in the upper left of the panels. 
For these two AGNs, NGC 5548 and NGC 4151, the mean timescales  of four kinds of $f$ are almost the  same within errors,  $638, 668~\rm days$, respectively. 
Through simulations, it is suggested that the timescale can be obtained when the ratio baseline $T$ to  their variation timescale $\tau_{\rm D}$ is larger than 10 \citep{Kozlowski_2017, Hu_2024}, although  other results suggest $T/\tau_D > 3$ and $30$  \citep{Kozlowski_2021, Suberlak_2021}. T=11481 and 10467 for NGC 5548 and NGC 4151, respectively,  which are larger than 10 times of  $\tau_{\rm D}$ of around 600 days. At the same time,  $\tau_{\rm D}$ is close to or larger than  the average cadence of  about 522 and 698 days for NGC 5548 and NGC 4151. The DRW model can effectively describe the variations of virial factor for the long baseline for these two AGNs and give reliable $\tau_{\rm D}$.

BLRs dynamical timescale $\tau_{\rm dyn} \sim R/v$, where $R$ is the distance from BLRs to the SMBH and $v$ is represented by the Kepler velocity. Here, we use $\rm FWHM_{mean}$ as $v$, and use  the \hb lag  to calculate $R$. For both targets, $\tau_{\rm dyn}\sim 500~{\rm light~days}$, which is consistent with the DRW damping timescale $\tau_{\rm D} \sim 600~{\rm light~days}$ considering uncertainties. 
This indicates that the variation in virial factor $f$ is intrinsically associated with BLRs dynamical processes. \citet{Wang_2025} also indicates that for some high-luminosity AGNs, BLR kinematics can change over relatively short timescales (hundreds of days).
It is possible that  some BLRs dynamical processes lead to the variation of the BLRs, showing the variation of BLR geometry and inclination. 

\subsection{The UV-Optical time lag} \label{subsec:4.3}
\begin{figure*}
    \centering
    \subfigure{\label{Fig:R81}
    \includegraphics[angle=0,width=0.8\textwidth]{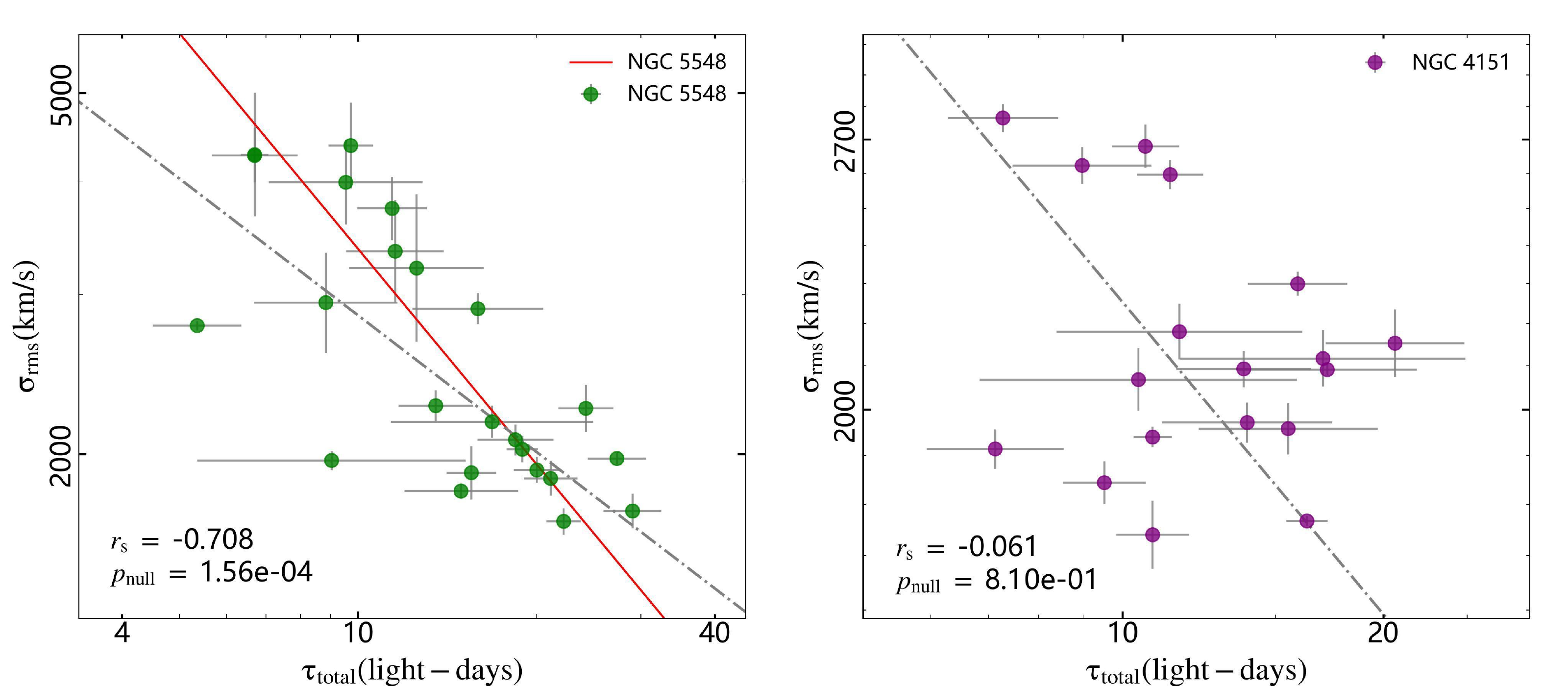}}
    \caption{Correlations between $\sr$ and the total time lags.  Symbols and  lines  are the same as which in Fig~\ref{Fig:1}.}.
    \label{Fig:8}
\end{figure*}  

\begin{figure*}
	\centering
	\subfigure{\label{Fig:R91}
	\includegraphics[angle=0,width=0.8\textwidth]{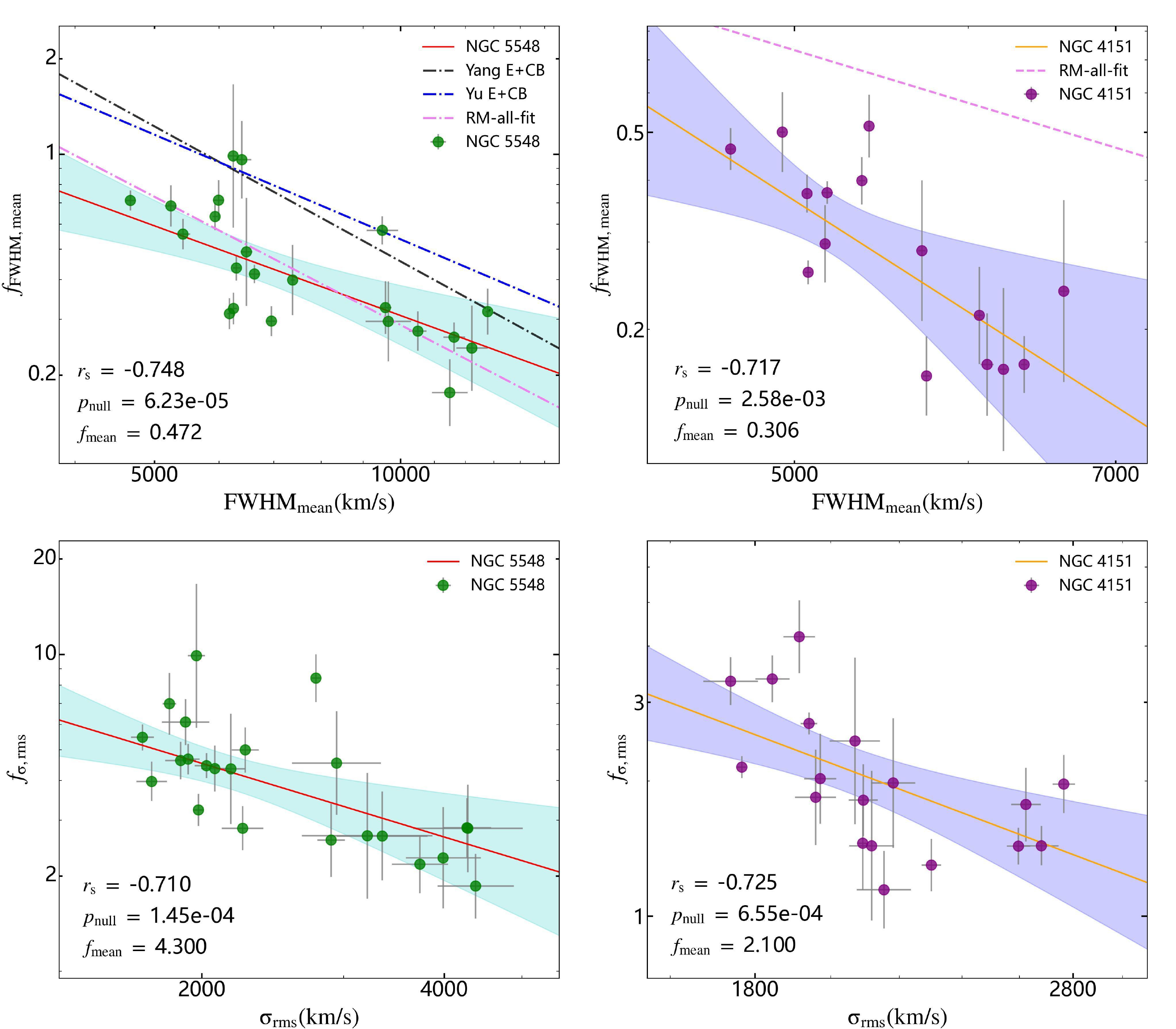}}
	\caption{Correlations between $f_\text{FWHM,mean}$, $\fsr$ considering the UV-Optical time delay and the corresponding broad line widths, respectively. Symbols, lines and the shaded areas are the same as which in Fig \ref{Fig:3}.}.
	\label{Fig:9}
\end{figure*}  

\begin{deluxetable*}{ccccccccc}
\tablecolumns{9}
\renewcommand\arraystretch{1.382}
\tablewidth{0pt}
\tablecaption{Linear regression results between the H${\beta}$ broad line widths and the total time lags for NGC 5548 and NGC 4151.
\label{tab:6}}
\tablehead{
& \multicolumn{4}{c}{log$_{10} \tau$ (NGC 5548)} &\multicolumn{4}{c}{log$_{10} \tau$ (NGC 4151)} \\
    \cmidrule(lr){2-5} \cmidrule(lr){6-9} 
     & $\alpha$ & $\beta$ &$r_\text{s}$ & $p_\text{null}$ & $\alpha$ & $\beta$ &$r_\text{s}$ & $p_\text{null}$ 
}
\startdata
log$_{10}$F$_\text{m}$ & -0.77$\pm$0.05 & 4.77$\pm$0.06 & -0.58 & 0.01 & - & - & 0.46 & 0.07  \\
log$_{10}$F$_\text{r}$ & -0.75$\pm$0.06  & 4.49$\pm$0.06  & -0.71  & 0.00 & - & - & 0.03 & 0.92  \\
log$_{10}$$\sigma_{\text{m}}$ & -0.66$\pm$0.04 & 4.12$\pm$0.08 & -0.71 & 0.00 & - & - & 0.13 & 0.64  \\
log$_{10}$$\sigma_{\text{r}}$ & -0.79$\pm$0.08 & 4.32$\pm$0.10  & -0.71 & 0.00 & - & - & -0.06 & 0.81  \\
\enddata
\tablecomments{\footnotesize $\alpha$ and $\beta$ respectively represent the slope and the intercept of the regressions. }
\end{deluxetable*}

\begin{deluxetable*}{cccccc}
\tablecolumns{6}
\renewcommand\arraystretch{1.382}
\tablecaption{Linear regression results between Virial Factors $f$ considering the total time lags and corresponding velocities for NGC 5548 and NGC 4151.
 \label{tab:7}}
\tablehead{
\colhead{}                 &
\colhead{}                  &
\colhead{$\alpha$}                   &
\colhead{$\beta$}                       &
\colhead{$r_{\rm s}$}                   &
\colhead{$p_{\rm null}$} 
}
\startdata
\hline
& log$_{10}$f$_\text{F,m}$ & -0.95$\pm$0.10 & 3.27$\pm$0.38 & -0.75 & $6.23\times 10^{-5}$  \\
NGC 5548 & log$_{10}$f$_\text{F,r}$ & -0.88$\pm$0.15  & 3.08$\pm$0.56  & -0.68  & $7.57\times 10^{-4}$  \\
& log$_{10}$f$_{\sigma,\text{m}}$ & -0.59$\pm$0.11 & 2.53$\pm$0.35 & -0.66 &$5.64\times 10^{-4}$  \\
&log$_{10}$f$_{\sigma,\text{r}}$ & -0.77$\pm$0.14 & 3.21$\pm$0.47 & -0.71 &$1.45\times 10^{-4}$ \\
\hline
& log$_{10}$f$_\text{F,m}$ & -2.84$\pm$0.41 & 10.06$\pm$1.53 & -0.72 & $2.58\times 10^{-3}$ \\
NGC 4151& log$_{10}$f$_\text{F,r}$ & -1.78$\pm$0.13 & 6.15$\pm$0.47 & -0.72 &$6.91\times 10^{-4}$  \\
 & log$_{10}$f$_{\sigma,\text{m}}$& - & - & -0.48 &   $7.11\times 10^{-2}$  \\
&log$_{10}$f$_{\sigma,\text{r}}$ & -1.40$\pm$0.18 & 4.96$\pm$0.61 & -0.73 &  $6.55\times 10^{-4}$  \\
\hline
\enddata
\tablecomments{\footnotesize $\alpha$ and $\beta$ respectively represent the slope and the intercept of the regressions. }
\end{deluxetable*}

In the virial \mbh Equation~\ref{eq:1}, $R_{\rm BLR}$ is the distance from the BLRs gas to the SMBH. Through \hb RM method, $\tau_{\rm \hb}$ only gives the distance between the broad \hb and the location in accretion disk emitting $L_{\rm 5100}$, and an additional distance from the accretion disk location to the SMBH should also be taken into account, which can be estimated through the UV-optical lag,  $\tau_{\rm X-5100}$ \citep{Fausnaugh_2016, Edelson_2017, Feng_2024, Zhou_2025}. 
From the continuum RM Swift UVOT  monitoring, $\tau_{\rm X-5100} = 2.58~ {\rm light~ days}$ for NGC 5548 \citep{Fausnaugh_2016, Kamoun_2021}. For NGC 4151, $\tau_{\rm X-5100} = 1.25\pm 0.40 ~ {\rm light~ days}$ in low \lv  with Swift UVOT  monitoring and $4.78\pm 0.63 ~ {\rm light~ days}$ in high \lv  with high-cadence spectroscopy monitoring, suggesting a variable additional UV-optical time lag depending on \lv \citep{Edelson_2017,  Feng_2024, Zhou_2025}. Considering large uncertainties on the relation between  $\tau_{\rm X-5100}$ and \lv, we adopt $\tau_{\rm X-5100}=4.02 ~ {\rm light~ days}$ for NGC 4151.

Fig~\ref{Fig:8} show the correlations between the total time lag $\tau_{\rm total}$ and $\sigma_{\rm rms}$. 
For NGC 5548, $r_{\rm s}=-0.71,~ p_{\rm null}=1.56 \times 10^{-4}$, and the best fitting gives $\sigma_{\rm rms} \propto \tau_{\rm total}^{-0.79\pm0.08}$, the slope is smaller than $-0.66\pm 0.07$ for $\thb$ without considering additional UV-optical time lags.
For NGC 5548, the slope deviation from the virial relation of  $-0.5$  becomes severe, and the same for other velocity tracers. 
For NGC 4151, there is still no obvious correlation between $\tau_{\rm total}$ and $\sigma_{\rm rms}$ with $r_{\rm s}=-0.06$ and  $p_{\rm null}=0.81$. The fitting results are shown in Table ~\ref{tab:6}. 

Considering additional  UV-optical time lags, four kinds of corrected $f$ all become smaller. Fig \ref{Fig:9} gives the relations between the corrected $\ffm$, $\fsr$ and corresponding broad line widths. 
For NGC 5548, the correlation between the corrected $f_{\rm FWHM,mean}$ and FWHM$_{\rm mean}$ becomes stronger with $r_{\rm s}$ from $-0.59$ to $-0.75$. The best fitting shows $\ffm \propto {\rm FWHM}_{\rm mean}^{-0.95\pm0.10}$. 
For NGC 4151, $r_{\rm s}$ is from $-0.69$ to $-0.74$. The best fitting shows $\ffm \propto {\rm FWHM}_{\rm mean}^{-2.90\pm0.42}$.
For the $\fsr-\sr$ relation, the absolute $r_{\rm s}$ are larger than 0.7 for both targets, with $p_{\rm null}$ both smaller than $1\times10^{-3}$. 
For the other velocities tracers, the correlations also become stronger. 
Therefore, including additional UV-optical lag strengthens the relations between virial factor and four velocity tracers.
The fitting results are shown in Table ~\ref{tab:7}. 

We also investigate the relations between $f$ and other observational parameters(i.e., \ledd, \dhb), no significant differences were found comparing with the results from $f$ calculated with $\thb$ without considering additional UV-optical time lags.


\section{Conclusions}\label{sec:5}

NGC 5548 and NGC 4151 had been spectral RM monitored over 20 observing seasons in the  past 30 years with negligible growth of central black hole mass during RM campaigns. For these two AGNs with the largest number of  spectral RM, we calculate the virial factor $f$ from the broad H${\beta}$ line width, \hb lag and \mbh from the BLRs dynamical model to investigate the variation of $f$ and related BLRs property. The main conclusions are summarized as follows:

\begin{enumerate}
\item For NGC 5548, correlations are found between $\thb$ and all four velocity tracers ($v\sim\thb^{\alpha},~\alpha:-0.55--0.66$), which has a deviation from the virial relation. For NGC 4151, no correlations were found for $v-\thb$. $f$ for this two targets are similar, and the ranges are given as follows, log$\ffm$: $-0.8\sim0.2$; log$\ffr$: $-0.7\sim0.6$; log$\fsm$: $0.1\sim1.1$; log$\fsr$: $0.1\sim1.2$. Each kind of $f$ has a range of about one order of magnitude.
    
\item It is found there are strong negative correlations between $f_{\rm F,mean}$ and  FWHM$_{\rm mean}$, for NGC 5548, $f_{\rm F,mean}\propto  {\rm FWHM}_{\rm mean}^{-0.70\pm0.13}$ and for NGC 4151, $f_{\rm F,mean}\propto {\rm FWHM}_{\rm mean}^{-3.31\pm0.59}$.  
Similar relations also exist for $\Fr$. For $f_{\sigma,\text{rms}}-\sr$, a relative strong relation appears, with $r_{\rm s}=-0.57$ for NGC 5548 and -0.47 for NGC 4151, 
while for $\fsm-\sm$, the relation becomes weaker or disappear for both targets. It suggests that the variable $f$ should also be considered when calculating virial \mbh with FWHM or $f_{\sigma,\text{rms}}$ for an individual AGN in  different RM epochs.
    
\item For NGC 5548 and NGC 4151, moderate correlations are found between $f_{\rm F,rms}$ and $\leddR$ ($r_{\rm s}=-0.59, -0.51$ respectively); $f_{\rm F,mean}$ and $\dhbm$ ($r_{\rm s}=-0.52, -0.56$ respectively). 
For NGC 5548,  there is a moderate correlation between $f_{\sigma,{\rm rms}}$ and $L_{\rm Bol}/L_{\rm Edd}$ ($r_{\rm s}=-0.55$) .
For NGC 4151, there is a strong anti-correlations between $f_{\rm FWHM,rms}$ and $D_{\rm \hb,rms}$  ( $r_{\rm s}= -0.75$).
Relations between $f$ and the line profile shape show that  $f$ is associated with the dynamics of BLRs.

    
\item Using a simple model of thick-disc BLRs, we constrain the BLRs inclination $\theta$ and the scale height of the thick disk $a$ for both targets through fitting  the cumulative proportion of  four kinds of $f$. For NGC 5548, $\theta_{\rm min}=14.1\pm 4.5$ deg, $\theta_{\rm max}=40.6\pm 2.0$ deg, $a=0.18\pm 0.10$. For NGC 4151, $\theta_{\rm min}=14.0\pm 4.0$ deg, $\theta_{\rm max}=55.1 \pm 2.7$ deg, $a=0.13 \pm 0.09$. $\theta_{\max}$ for NGC 5548 is $\sim$15 deg less than  NGC 4151. It indicates that the inclination variation  leads to the variation of $f$ for an individual AGN.

\item Considering additional  UV-Optical time lags, the deviation from the virial relation becomes more severe for NGC 5548 ($v\sim\tau^{\alpha},~\alpha:-0.66--0.79$), while still no relations are found for NGC 4151. The correlation between $f_{\rm F,mean}$ and FWHM$_\text{mean}$ strengthens with $r_{\rm s}$ reaching $-0.75$ for NGC 5548 and $-0.72$ for NGC 4151. 
This indicates that including additional UV-optical lag leads to stronger relations between the corresponding $f$ and four velocity tracers. 

\item Using the DRW model, the damping timescale of four kinds of $f$ for both targets is $\tau_{\rm D}\sim 600$ light days on average, which is consistent with the dynamical timescale $\tau_{\rm dyn}\sim$ 500 light days within uncertainties. This indicates that the variations in virial factor $f$ are intrinsically associated with BLRs dynamical processes, leading to the variation of the BLRs and showing the variation of BLR geometry and inclination.

\end{enumerate}


\begin{acknowledgments}
We are  very grateful to the anonymous referee for her/his instructive comments which significantly improved the content of the paper. 
This work has been supported by the National Science Foundations of China (No. 11973029 ).
\end{acknowledgments}

\bibliographystyle{aasjournal}
\bibliography{vf_bian_arxiv}


\end{document}